\documentclass[10pt]{iopart}
\usepackage{iopams}
\usepackage{setstack}
\usepackage{amssymb}
\usepackage[square,sort&compress]{natbib}
\usepackage{graphicx}

\begin{document}

\title[Relativistic Astrometry for a Gaia-like observer]{The Ray Tracing Analytical Solution within the RAMOD framework. The case of a Gaia-like observer.}

\author{M Crosta$^1$, A Vecchiato$^1$ F de Felice$^2$, M G Lattanzi$^1$}

\address{$^1$  Astrophysical Observatory of Torino, INAF, via Osservatorio 20, I-10025 Pino Torinese (TO), Italy}
\address{$^2$ Department of Physics and Astronomy, University of Padova, via Marzolo 8, I-35131 Padova, Italy}
\ead{crosta@oato.inaf.it}
\vspace{10pt}
\begin{indented}
\item[]October 2014
\end{indented}

\begin{abstract}
This paper presents the analytical solution of the inverse ray tracing problem for photons emitted by a star and collected by an observer located in the gravitational field of the Solar System. This solution has been conceived to suit the accuracy achievable by the ESA Gaia satellite (launched on December 19, 2013) consistently with the measurement protocol in General relativity adopted within the RAMOD framework. Aim of this study is to provide a general relativistic tool for the science exploitation of such a revolutionary mission, whose main goal is to trace back star directions from within our local curved space-time, therefore providing a three-dimensional map of our Galaxy. The results are useful for a thorough comparison and cross-checking validation of what already exists in the field of Relativistic Astrometry.
Moreover, the analytical solutions presented here can be extended to model other measurements that require the same order of accuracy expected for Gaia.
\end{abstract}
%\noindent{\it Keywords\/ general relativity; gravitation; astrometry; analytical methods; Galaxy}
\submitto{\CQG}

\maketitle
%\ioptwocol

\section{Introduction}
To fully exploit the science of the Gaia mission (ESA, \citep{2005tdug.conf.....T}), a relativistic  astrometric model is needed able to cope with an accuracy of  few $\;\mu\mathrm{as}$ for observations within the Solar System.

Gaia acts as a celestial compass, measuring arches among stars with the purpose to determine their position via the absolute parallax method. The main goal is to construct a three-dimensional map of the Milky Way and unravel its structure, dynamics, and evolutional history. This task is accomplished through a complete census, to a given brightness limit, of about one billion individual stellar objects.

Since the satellite is positioned at Lagrangian point L2 of the Sun-Earth system, the measurements of Gaia are performed in a weak gravitational regime and the solution of Einstein equation, i.e the space-time metric, has the general form
\begin{equation}
g_{\alpha\beta}=\eta_{\alpha\beta}+h_{\alpha\beta} +\mathcal{O}\left(h^{2}\right),
\label{metric}
\end{equation}
where $|h_{\alpha\beta}|\ll1$ and $|\partial_i h_{\alpha\beta}|\ll1$ can be treated as perturbations of a flat space-time and represent all the Solar System contributions to the gravitational field.
Their explicit expression, however, can be described in different ways according to the physical situation we are considering. This means that, for the weak-field case, $h_{\alpha\beta}$ can always be expanded in powers of a given smallness parameter $\epsilon$, as
\[
h_{\alpha\beta}=\underset{(1)}{h_{\alpha\beta}}+\underset{(2)}{h_{\alpha\beta}}+\underset{(3)}{h_{\alpha\beta}}+\cdots,\]
where the underscript  $n$ indicates the order of $\epsilon$. This expansion is usually made in power of the gravitational constant {\it G} (post-Minkowskian approach) or in power of $1/c$ (post-Newtonian approach) both approaches coinciding inside the near zone of the Solar System \citep{1987thyg.book.....H}.  While the post-Minkowskian formalism is better suitable outside the near-zone of the Solar System, the estimates performed inside this zone are sufficiently well supported by an approximation to the required order in $(v/c)\sim \epsilon$  which amounts to about $10^{-4}$ for the typical velocities of our planet.  Moreover, for the propagation of the light inside the Solar System, the sources of gravity should be considered  together with their internal structure and geometrical shape. This is particularly true when the light passes close to the giant planets. In other circumstances it is an unnecessary complication to consider the planets different from point-like objects especially when the model is devoted to the reconstruction of the \emph{stellar positions} in a global sense.  At the microarcsecond level of accuracy, {\it  i.e.} $(v/c)^3\sim \epsilon^3$, the contribution to the metric coefficients of the motion and the internal structure of the giant planets should be taken into account, in particular if one wants to measure specific light deflection effects, as for example,  those due to the quadrupolar terms. 

The scope of this paper is to present an analytical solution for a null geodesic of the metric (\ref{metric}) consistently with the requirements of Gaia's astrometric mission and according to the RAMOD framework \citep{2004ApJ...607..580D, 2006ApJ...653.1552D}. 
RAMOD uses a 3+1 description of the space-time in order to measure physical effects along the proper time and in the rest-space of a set of fiducial observers according to the following measurement protocol \citep{2010ToM.book.....D}: 

i) specify the phenomenon under investigation; 

ii) identify the covariant equations which describe the above effect; 

iii) identify the observer who makes the measurements;

iv) chose a frame adapted to that observer allowing the space-time splitting into the observer's space and time; 

v) understand the locality properties of the measurement under consideration (namely whether it is local or non-local with respect to the background curvature); 

vi) identify the frame components of the quantities which are the observational targets; 

vii) find a physical interpretation of the above components following a suitable criterium; 

viii) verify the degree of the residual ambiguity, if any, in the interpretation of the measurements and decide the strategy to evaluate it (i.e. comparing to what already known).

The main procedure of the RAMOD approach is to express the null geodesic in terms of the physical quantities which enter the process of measurement, in order to entangle the entire light trajectory with the background geometry at the required approximations. Then, the solution is adapted to the relevant IAU resolutions considered for Gaia \cite{2010A&A...509A..37C}. 

Solving the astrometric problem turns out to compile an astrometric catalog  with the accuracy of the measurement model.
Indeed there exist several models conceived for the above task and formulated in different and independent ways (\citep{2003AJ....125.1580K, 2012CQGra..29x5010T, 2014PhRvD..89f4045H} and references therein). Their availability must not be considered as an ``oversized toolbox'' provided by the theoretical physicists. Quite the contrary, they are needed to put the future experimental results on solid grounds, especially if one needs to implement gravitational source velocities and retarded time effects. From the experimental point of view, in fact, modern space astrometry is going to cast our knowledge into a widely unknown territory. Such a huge push-forward will not only come from high-precision measurements, which call for a suitable relativistic modeling, but also in form of  \emph{absolute} results which need be validated by independent, ground-based observations. In this regard, it is of capital importance to have  \emph{different}, and  \emph{cross-checked} models which exploit different  \emph{solutions} to interpret these experimental data. 

For the reason above, inside the Consortium  constituted for the Gaia data reduction (Gaia CU3, Core Processing, DPAC), two models have been developed: i) GREM (Gaia RElativistic Model, \cite{2003AJ....125.1580K}) baselined for the Astrometric Global Iterative Solution for Gaia (AGIS),  and ii) RAMOD (Relativistic Astrometric MODel) implemented in the Global Sphere Reconstruction (GSR) of the Astrometric Verification Unit at the Italian data center (DPCT, the only system, together with the DPC of Madrid, able to perform the calibration of positions, parallaxes and proper motions of the Gaia data).
RAMOD was originated to satisfy the validation requirement and, indeed, the procedure developed  can be conceived  to all physical measurements which imply light propagation.

Section \ref{sec2} lists the notation used in this paper; section \ref{sec3} is devoted  to the definition of the mathematical environment needed to make the null geodesic explicit at the desired accuracy. In particular,  in order to fully accomplish the precepts of the measurement protocol and to isolate the contributions from the derivative of the metric terms at the different retained orders, we introduce a suitable classification of  the RAMOD  equations. In section \ref{sec4} we set the appropriate approximations which permit the analytical solution of the astrometric problem. In section \ref{sec5}, we show the specific solution for the light deflection by spherical and non-spherical gravitational sources.  In section \ref{sec6}, finally, we deduce the analytical solution of the trajectories of the light signal emitted by the stars and propagating through the gravitational field within the Solar System.    
In the last section we summarize the conclusions.

\section{Notations}
\label{sec2}
\begin{itemize}
\item Greek indeces run from 0 to 3, whereas Latin indeces from 1 to 3;
\item "$_{,\alpha}$":  partial derivative with respect to the $\alpha$ coordinate;
\item"$\cdot$": scalar product with respect to the  euclidean metric $\delta_{ij}$;
\item"$\times$": cross product with respect to the euclidean metric $\delta_{ij}$;
\item tilde-ed symbols "$\,\, \tilde{}\,\,$" refer to quantities related to the gravitational sources;
\item repeated indeces like $ u_{\alpha} v^{\alpha}$ for any four vectors  $u^{\alpha}, v^{\alpha}$ means summation over their range of values;
\item "$[ \alpha \beta]"$: antisymmetrization of the indeces  $ \alpha, \beta$
\item $\mathbf{\nabla}$: covariant derivative;
\item  $\nabla_{\alpha}$: $\alpha$-component of the covariant derivative;
\item  $\nabla (f)$:  spatial gradient of a function $f$                                                                                                                                                                                                                                                                                
\item we use geometrized unit, namely $G=1$ and, $c=1$.
\end{itemize}

\section{The RAMOD equations for Gaia}
\label{sec3}
The basic unknown of the RAMOD method is the space-like four-vector $\bar{\ell}^{\alpha}$ , which is the projection of the tangent to the null geodesic into the rest-space of the local barycentric observer, namely the one locally at rest with respect to the barycenter of the Solar System. Physically, such a four-vector identifies the \emph{line of sight} of the incoming photon relative to that observer.

Once defined $\bar{\ell}^{\alpha}$,  the equations of the null geodesic takes a form which we shall refer to as master-equations. Neglecting all the $O(h^{2})$ terms, these read \citep{2006ApJ...653.1552D, 2011CQGra..28w5013C}:
\begin{eqnarray}
\frac{d\bar\ell^0}{d\sigma}&-&\bar\ell^i\bar\ell^j h_{0j,i}-\frac{1}{2}  h_{00,0}=0, \label{eq:me0}\\
&{}&\nonumber\\
\frac{d\bar\ell^k}{d\sigma}&-&\frac{1}{2} \bar\ell^k\bar\ell^i (\bar\ell^j h_{ij,0} - h_{00,i}) +\bar\ell^i\bar\ell^j\left( h_{kj,i}-\frac{1}{2} h_{ij,k}\right) 
\label{eq:mek} \\
&+& \bar\ell^i\left(h_{k0,i}+h_{ki,0} - h_{0i,k}\right) -\frac{1}{2} h_{00,k} -\bar \ell^{k} \bar \ell^{i} h_{0i, 0} + h_{k0,0} =0. \nonumber 
\end{eqnarray}
Here $\sigma$ is the affine parameter of the geodesic and 
\begin{equation}
\mathrm{d}\sigma=\mathrm{d}t+\mathcal{O}\left(h\right),\label{eq:reldsdt}
\end{equation}
whereas $t$ is the coordinate time.

In order to solve for the master equations one should define appropriate metric coefficients. To the order of $\epsilon^{3}$,  which is what is required for the accuracy targeted for Gaia, one has to take into account the distance between the points on the photon trajectory and the barycenter of the a-th gravity source at the appropriate retarded time together with the dynamical contribution to the background metric by the relative motion of the gravitational sources.  
More specifically ${r}_{(a)}^i$ is the retarded distance defined as  
\begin{equation}
r^i_{(a)} (\sigma, \tilde \sigma') =  x^i(\sigma) -\tilde x_a^i (\tilde \sigma'),
\label{eq:retdist}
\end{equation}
where $\tilde \sigma$ is the parameter of the $(a)$-th source's world line. The retarded
position of the source is fixed by the intercept of its worldline with the past light cone at any point on the photon trajectory.  However, the retarded time $t'=t-r_{(a)}$ and the retarded distance $r_{(a)}$ are intertwined in an implicit relation which would prevent us to solve the geodesic equations. Nonetheless,  we show that it is possible to write an approximate form of the metric which retains the required order of accuracy of $\epsilon^3$, but where the dependence from the retarded contribution is simplified. 
 
By  using the Taylor expansion around any $\tilde \sigma'$ to the first order in $\epsilon$, we get for each source:
\begin{equation}
 x^i (\tilde \sigma) \approx  \tilde{x}^i (\tilde \sigma' ) + \tilde v^i (\tilde \sigma') (\tilde \sigma-\tilde \sigma'), 
\end{equation}
which allows to rewrite the retarted distance as 
$$r^i_{(a)}= x^i(\sigma) - \tilde x^i(\tilde \sigma') \approx  x^i(\sigma) - \tilde x^i(\tilde \sigma)+ \tilde v^i (\tilde \sigma') (\tilde \sigma' -\tilde \sigma) ,$$ 
i.e.
\begin{equation}
 r_{(a)}^i= r^i  (\sigma, \tilde \sigma)+\tilde r^i (\tilde \sigma, \tilde \sigma') + O(\tilde v^2), 
\end{equation}
where we set $ \tilde r^i (\tilde \sigma, \tilde \sigma') = \tilde x^i (\tilde \sigma) -  \tilde x^i (\tilde \sigma')$.
Nevertheless $(\tilde \sigma' -\tilde \sigma)$ is again proportional to the retarded distance as measured along the source world line. In fact, considering the tangent four-vector of the source world line
$$\tilde u^{\alpha} = - (u_{\beta} \tilde{u}^{\beta})  (u^{\alpha}  + \tilde v^{\alpha}),$$ where  $\tilde v^{\alpha}$ is the $\alpha$-component of the spatial four-velocity of the source relative to the origin of the coordinate system and defined in the rest frame of the local barycentric observer $\mathbf{u}$, the interval elapsed from the position of the source at the time $t'$ and that at $t$ is
\begin{eqnarray}
\tilde \sigma'- \tilde \sigma &=&  \Delta T_{\tilde u} = - \tilde u_{\alpha} \Delta x^{\alpha} \nonumber \\
&=& -\eta_{\alpha \beta} \tilde u^{\alpha} \Delta x^{\beta} + O(h) \nonumber \\
&\approx& \Delta x^0 - \delta_{ij} \tilde v^i \Delta x^j +  O(h), 
\label{eq:ret_time_pla} 
\end{eqnarray}
where $\Delta x^{\alpha}= x^{\alpha}(\tilde \sigma (t')) -x^{\alpha}(\tilde \sigma (t))$.
To the first order in $\tilde v$, we have along the generator of the light cone 
\begin{equation}
 \Delta x^0 =  r(\sigma, \tilde \sigma)+ \frac{(\mathbf{r}(\sigma, \tilde \sigma) \cdot \mathbf{\tilde r} (\tilde \sigma, \tilde \sigma') )}{ r(\sigma, \tilde \sigma)} +  O(\tilde v^2),
\end{equation}
then we get the following approximate expression for (\ref{eq:retdist}):
\begin{equation}
 r^i(\sigma, \tilde \sigma')_{(a)} \approx  r^i  (\sigma, \tilde \sigma) -   \tilde v^i(\tilde \sigma')  r (\sigma, \tilde \sigma) 
 \label{eq:final_ret_dis}
\end{equation}
or
\begin{equation}
r^i(\sigma, \tilde \sigma')_{(a)}  \approx r  (\sigma, \tilde \sigma) [ n^i  (\sigma, \tilde \sigma) -   \tilde v^i (\tilde \sigma')], 
\label{eq:final_ret_dis}
\end{equation}
where $ n^i  (\sigma, \tilde \sigma) =  r^i /r$. This is equivalent, to first order in $\tilde v$, to the distance found in \cite{1999PhRvD..60l4002K} and entering  the expression of the metric, i.e.
\begin{equation}
r (\sigma, \tilde \sigma')_{(a)}  \approx r   -  \mathbf{r}\cdot \mathbf{\tilde v}. 
\label{eq:final_ret_dis_kopetsch}
\end{equation}

 The choice for the perturbation term of the metric has to match the adopted retarded distance approximation and the fact that the lowest order of the $h$ terms is $\epsilon^2$ and the present space astrometry accuracy does not exceed the $\epsilon^3$ level.

Then, for our purpose, a standard suitable solution of Einstein's equations  in terms of
a retarded tensor potential \citep{1990recm.book.....D,2006ApJ...653.1552D}, which can be further specialized as the Li\'enard-Wiechert potentials \citep{1999PhRvD..60l4002K} is  
\begin{eqnarray}
h_{00} & = & \sum_{a}\frac{2\mathcal{M}_{(a)}}{r_{(a)}} +\mathcal{O}\left(\epsilon^4 \right)\nonumber \\
h_{0i} & = &-\sum_{a}\frac{4\mathcal{M}_{(a)}}{r_{(a)}}\tilde \beta_{i(a)}+\mathcal{O}\left(\epsilon^{5 }\right)\label{eq:metric-explicit}\\
h_{ij} & = & \sum_{a}\frac{2\mathcal{M}_{(a)}}{r_{(a)}}\delta_{ij}+\mathcal{O}\left( \epsilon^4 \right),\nonumber \end{eqnarray}
where $\mathcal{M}_{(a)}$ is the mass of the $a$th gravity source,
$\mathbf{r}_{(a)}$ is the position vector of the photon with respect
to the source, $\tilde \beta^j = \tilde x^j_{,0} = (1-h_{00}/2) \tilde{v}^i(\tilde \sigma) + O(h^2) $ is the coordinate spatial velocity of the gravity source. 

Note that the time component of the tangent vector to the source's worldline \citep{2006ApJ...653.1552D} is
\begin{equation}
\tilde u^0 = \frac{dt}{d \tilde \sigma}= 1 + \frac{h_{00}}{2} + \frac{\tilde v^2}{2c^2 }+ O (h^2) +O\left(\epsilon^4 \right)
\label{eqdsigmau0}
\end{equation}
while that of the local barycentric observer is
\begin{equation}
 u^0 = \frac{ dt}{d  \sigma}= 1 + \frac{h_{00}}{2} + O (h^2).
 \label{eqdsigmau'0}
\end{equation}
Then from (\ref{eqdsigmau0}) and (\ref{eqdsigmau'0}) we derive the following relationships in the linear approximation
\begin{equation}
d \tilde \sigma=  dt \left( 1 - \frac{h_{00}}{2} - \frac{\tilde v^2}{2} \right) + O (h^2) +O\left(\epsilon^4\right),
\end{equation}
and
\begin{equation}
d \sigma=  dt \left( 1 - \frac{h_{00}}{2}  \right) + O (h^2) +O\left(\epsilon^4\right).
\end{equation}

Within the approximation (\ref{eq:final_ret_dis_kopetsch}) the perturbation of the metric transforms as
\begin{eqnarray}
h_{00} & = & 2\sum_{a}\frac{\mathcal{M}_{(a)}}{r_{(a)}} (1+\mathbf{n}_{(a)}\cdot \mathbf{\tilde v}_{(a)} ) +\mathcal{O}\left(\epsilon^4 \right)\nonumber \\
h_{0i} & = & - 4  \sum_{a}\frac{\mathcal{M}_{(a)}}{r_{(a)}}v_{i(a)}+\mathcal{O}\left(\epsilon^{5 }\right)\label{eq:metric-explicit-firstorderv}\\
h_{ij} & =  & 2 \sum_{a}\frac{\mathcal{M}_{(a)}}{r_{(a)}}(1+\mathbf{n}_{(a)}\cdot \mathbf{\tilde v}_{(a)} ) \delta_{ij}+\mathcal{O}\left( \epsilon^4 \right)
\nonumber \end{eqnarray}
or, by simplifing the notation
\begin{eqnarray}
 h_{00} & =&  h   \simeq  \sum_{a}  h_{(a)} = 2 \sum_{a}  \mathcal{M}_{(a)} \frac {1}{r_{(a)}} (1+\mathbf{n}_{(a)}\cdot\mathbf{\tilde v}_{(a)} )+  \mathcal{O}(\epsilon^{4})  \nonumber \\
  h_{0i} &  =&  - 2  h \tilde v_{i}  \simeq -2 \sum_{a}  h_{(a)}  \tilde v_{i(a)} \label{eq:me-simplify} +  \mathcal{O}(\epsilon^{3})  \label{eq:me-simplify} \\
 h_{ij}  & \simeq & h\,\delta_{ij} + \mathcal{O}(\epsilon^{4}), \nonumber 
\end{eqnarray}

In the follows, $r_{(a)}$, unless explicitly expressed,  will indicate a function  with arguments $\sigma$ and $\tilde \sigma$.
Moreover, to ease notation we drop the index (a) wherever it is not necessary.

\subsection{The n-bodies spherical case}
Let us consider a space-time splitting with respect to the congruence of fiducial observers $\mathbf{u}$ in the gravitational field of the Solar System \citep{2004ApJ...607..580D}. 
The field equations can  be rewritten in terms of the shear, expansion and vorticity of the congruence $\mathbf{u}$ 
 \citep[see][]{1990recm.book.....D}. For our purpose it is enough to consider only the expansion term and the vorticity \citep[see][]{2011CQGra..28w5013C}.

The master equations (\ref{eq:me0}) and (\ref{eq:mek}) are obtained  by retaining the vorticity term at least to the order of $O(h_{0i})$, and the expansion to the order of $O(\partial_{0}h_{0 0})$ and $O(\partial_{0}h_{0 i})$. 
In the case of a vorticity and expansion-free geometry, the RAMOD master equations are named RAMOD3 master equation \citep{2004ApJ...607..580D}
\begin{equation}
\fl
\frac{\mathrm{d}\bar{\ell}^{k}}{\mathrm{d}\sigma}+\bar{\ell}^{i}\bar{\ell}^{j}\left( h_{kj,i}- \frac{1}{2} h_{ij,k}\right)+\frac{1}{2}\bar{\ell}^{k} \bar{\ell}^{i}h_{00,i}- \frac{1}{2}h_{00,k}+\mathcal{O} \left(h^{2}\right)=0,
\label{eq:me-fv}
\end{equation}
where $\bar \ell^0=0$.
Taking into account that  $\bar{\ell}^{i}\bar{\ell}^{j}\delta_{ij}=1+\mathcal{O}\left(\epsilon^{2}\right)$, equations (\ref{eq:me0}), (\ref{eq:mek}), and (\ref{eq:me-fv}) can be reduced respectively to:
\begin{eqnarray}
\fl
\frac{d\bar\ell^0}{d\sigma} = -2   (\bar\ell \cdot \tilde v )( \bar\ell^i h_{,i}) + \frac{1}{2}  h_{,0}+  \mathcal{O}(h^{2})  \label{eq:diffeq0}\\
&{}&\nonumber\\
\fl
\frac{d\bar\ell^k}{d\sigma}= -\frac{3}{2}\bar{\ell}^{k}( \bar\ell^i h_{,i})  + h_{,k} -\frac{1}{2} \bar\ell^k  h_{,0} - 2   (\bar\ell \cdot \tilde v) h_{,k} + 2 \tilde v^k  ( \bar\ell^i h_{,i}) -   \bar{\ell}^{k} \bar\ell^i ( \tilde{v}_{i} h)_{,0}  \\
&{}&\nonumber\\
 +  ( \tilde{v}^k h)_{,0} + \mathcal{O}(h^{2})  
\label{eq:diffeqk}
\end{eqnarray}
and for the static case ($h_{\alpha \beta,0}=0$ and $\tilde v=0$)
\begin{equation}
\frac{\mathrm{d}\bar{\ell}^{k}}{\mathrm{d}\sigma} + \frac{3}{2}\bar{\ell}^{k}(\bar{\ell}^{i}h_{,i})- h_{,k}+\mathcal{O}(h^{2}) =0. \label{eq:master-equation-2}
\end{equation}
 
In order to fully accomplish the precepts of the measurement protocol, it would be useful, to isolate the contributions from the derivatives of the metric at the different retained orders.

This allows us to classify the master equation as follows:
\begin{itemize}
\item RAMOD3a (R3a), the spatial derivatives of the metric are considered while $h_{0i}$ are neglected
\begin{equation}
\frac{\mathrm{d}\bar{\ell}^{k}}{\mathrm{d}\sigma} = - \frac{3}{2}\bar{\ell}^{k}(\bar{\ell}^{i}h_{,i})+ h_{,k}+\mathcal{O}(h^{2})\equiv (\mathrm{R3a})^k  \label{eq:master-equation-r3a}
\end{equation}
\item RAMOD3b (R3b), the spatial and time derivatives of the metric are considered while $h_{0i}$ are neglected
\begin{eqnarray}
\frac{d\bar\ell^0}{d\sigma}&=&  \frac{1}{2}  h_{,0}+ \mathcal{O}(h^{2})\equiv (\mathrm{ R3b})^0   \label{eq:meq0-r3b}\\
&{}&\nonumber\\
\frac{d\bar\ell^k}{d\sigma}&=& (\mathrm{R3a})^k  - \frac{1}{2} \bar\ell^k  h_{,0}
 + \mathcal{O}(h^{2})\equiv(\mathrm{ R3b})^k
\label{eq:meqk-r3b}
\end{eqnarray}
\item RAMOD4a (R4a), the spatial derivatives of the metric are considered including $h_{0i}$
\begin{eqnarray}
\frac{d\bar\ell^0}{d\sigma}&=& (\mathrm{ R3b})^0  - 2  ( \bar\ell \cdot \tilde{v})( \bar\ell^i h_{,i}) + \mathcal{O}(h^{2})\equiv (\mathrm{ R4a})^0\label{eq:meq0-r4a}\\
&{}&\nonumber\\
\frac{d\bar\ell^k}{d\sigma}&= &(\mathrm{ R3b})^k - 2 ( \bar\ell \cdot \tilde{v}) h_{,k}  + 2 \tilde{v}^k  ( \bar\ell^i h_{,i}) \nonumber\\
&{}&\nonumber\\
&+& \mathcal{O}(h^{2})\equiv(\mathrm{ R4a})^k
\label{eq:meqk-r4a}
\end{eqnarray}
\item RAMOD4b (R4b),  the spatial and time derivatives of the metric are considered including $h_{0i}$
\begin{eqnarray}
\frac{d\bar\ell^0}{d\sigma}&=& (\mathrm{R4a})^0 + \mathcal{O}(h^{2}) \equiv (\mathrm{R4b})^0 \label{eq:meq0-r4b}\\
&{}&\nonumber\\
\frac{d\bar\ell^k}{d\sigma}&= &(\mathrm{ R4a})^k -2 \bar{\ell}^{k} \bar\ell^i ( \tilde{v}_{i} h)_{,0} +  2( \tilde{v}^k h)_{,0} + \mathcal{O}(h^{2})\equiv (\mathrm{R4b})^k
\label{eq:meqk-r4b}
\end{eqnarray}

\end{itemize}

The implementation of RAMOD models and the need of testing them through a self-consistency check at different levels of accuracy, will benefit form this explicit classification.    
Beside this new classification of the RAMOD master equations, it is clear that the solutions call for an explicit expression of the metric terms.
In general, for any integer $m$: 
\begin{equation}
 (r)^m _{,i}=  m  r^{i } (r)^{m-2}, \,\,\,\,\,\, (r)^m _{,0}=- m ( \tilde \beta \cdot r)  (r)^{m-2}.
\end {equation}

From the last computation one could expect that in the case of mapped trajectories for RAMOD3-like model \citep[see][]{2004ApJ...607..580D, 2011CQGra..28w5013C} the term $x^i_{,0}$ should be retained, since each mapped spatial coordinate depends on the $\sigma$ value of the local one-parameter diffeormorphism. In this respect, note that the null geodesic crosses each slice $S(t)$  at a point with coordinates $x^i(\sigma(t))$, but this point also belongs to the unique normal to the slice $S(t)$ crossing it with a value $\sigma=\sigma(x^i, t)$ which runs differently for any spatial coordinate and therefore does not coincide with the proper time of the local barycentric observer. Therefore  
$$\partial_0 x^i =( \partial_{\sigma} x^i(\sigma))( \partial_{0} \sigma(x^i(\lambda), t) ) =0.$$

Now, by using the retarted time approximation we get:
\begin{eqnarray}
\fl
(r)^{-m} _{,i}(\sigma, \tilde \sigma')& = &- m  r^{-m-1}_{(a)} (1+  n_{(a)} \cdot \tilde v )^{m-1} [ n^{i}_{(a)} (1+  2 n_{(a)} \cdot  \tilde v) - \tilde v^i] + O(h)
 \label{der-r-0exp}
\end {eqnarray}
and 
\begin{equation}
 (r)^{-m}_{,0} (\sigma, \tilde \sigma')=  m  r^{-m-1}_{(a)}  (n_{(a)} \cdot \tilde {\beta}) (1+ n_{(a)} \cdot \tilde {\beta})^m   + O(h).
\label{der-r-exp}
\end {equation}
  
Let us indicate the photon impact parameter with respect to the source position as:
\begin{equation}
d^k_{(a)}= r^k_{(a)}- \bar \ell^k (  r_{(a)} \cdot \bar \ell) 
\label{eq:sip-g}
\end{equation}
and, for sake of convenience, let us denote also:
\begin{equation}
d^k_{v_{(a)}}= \tilde v^k - \bar \ell^k (  \tilde v  \cdot \bar \ell). 
\label{eq:sip-g}
\end{equation}
Finally, according to the previous derivatives, making them explicit, and denoting $n^i= r^i/r$,  the master equations assume the following expressions, valid up to the $\epsilon^3$ order:
\begin{itemize}

\item RAMOD3a: 
\begin{eqnarray}
\fl
\frac{\mathrm{d}\bar{\ell}^{k}}{\mathrm{d}\sigma} = 2  \sum_{a}  \frac {\mathcal{M}_{(a)}}{r^2_{(a)}} \left\{ \left( \frac{1}{2 } \bar{\ell}^{k} (n_{(a)} \cdot \bar \ell )-  d^k \right) (1 + 2 n_{(a)} \cdot  \tilde v) -   \frac{1}{2 } \bar{\ell}^{k}( \bar\ell \cdot \tilde{v} ) +  d^k_{v}  \right\}  \nonumber \\
+  O(\tilde v^2)+  O(h^2)\label{eq:meq-r3af},
\end{eqnarray}
where in case of zero velocity we recover the static RAMOD recorded as follows

\item RAMOD3s(R3s):    
\begin{equation}
\fl
\frac{\mathrm{d}\bar{\ell}^{k}}{\mathrm{d}\sigma} = 2  \sum_{a}  \frac{\mathcal{M}_{(a)}} {r^2_{(a)}} \left\{ \frac{1}{2 } \bar{\ell}^{k} (n_{(a)} \cdot \bar \ell)- d^k \right\}  +  O(h^2)\label{eq:meq-r3s}.
\end{equation}

Similarly for the other classification items we have:

\item RAMOD3b:
\begin{eqnarray}
\fl
\frac{d\bar\ell^0}{d\sigma}=  \sum_{a}  \frac {\mathcal{M}_{(a)}} {r^2_{(a)}}   ( n_{(a)} \cdot \tilde v)    +  O(\tilde v^2)+  O(h^2) \label{eq:meq0-r3bf}\\
&{}&\nonumber\\
\fl
\frac{d\bar\ell^k}{d\sigma}=  (\mathrm{R3a})^k -  \bar\ell^k  \sum_{a}  \frac {\mathcal{M}_{(a)}} {r^2_{(a)}}   ( n_{(a)} \cdot \tilde v) +  O(\tilde v^2) +  O(h^2)
\label{eq:meqk-r3bf}
\end{eqnarray}
\item RAMOD4a
\begin{eqnarray}
\fl
\frac{d\bar\ell^0}{d\sigma}= (\mathrm{R3b})^0  + 4  \sum_{a}  \frac {\mathcal{M}_{(a)}} {r^2_{(a)}}  \left[ ( \bar\ell \cdot \tilde{v})( \bar\ell \cdot n_{(a)}) \right]  +  O(\tilde v^2)+  O(h^2) \label{eq:meq0-r4af}\\
&{}&\nonumber\\
\fl
\frac{d\bar\ell^k}{d\sigma}=(\mathrm{R3b})^k + 4  \sum_{a}  \frac {\mathcal{M}_{(a)}} {r^2_{(a)}}   \left[  \bar\ell \times  (n_{(a)} \times    \tilde{v}) \right]^k +  O(\tilde v^2)+  O(h^2)
\label{eq:meqk-r4af}
\end{eqnarray}

\end{itemize}

It is clear that, to the order of  $\epsilon^3$, we do not need to include the time derivative of $h_{0k}$ since these are at least of the order of $\epsilon^4$ and should be neglected. Therefore, we do not consider the part of the RAMOD4 equations which contain the time derivative of $h_{0k}$ and the second order velocity contributions.

\subsection{The case for an oblate body}

Now, let us consider the $a$-th source and define  
\begin{equation}
\fl
{h}_{(a)} =  2  M_{(a)} \bar h_{(a)} \equiv   2  M_{(a)}  \frac{1}{r_{(a)}}\left[ 1- \sum_{m=2}^{\infty}  J_m \left( 
\frac{ R_{(a) }} {r_{(a)}} \right)^m  P_m(\cos \theta_{(a)}) \right] \label{eq:hbar},
\end{equation}
which means to take into account the mass multipole structure of the a-th body where $P_m$ are the Legendre polynomials, $M_a$ the mass of the body, $ R_a$ its equatorial radius, $\theta_{(a)}$ the co-latitude, and  $J_m$  the coefficients of the mass multipole moments. With this choice our considerations are confined to the case in which the object' ellipsoid of inertia is an ellipsoid of revolution and the directions of the spatial coordinate axes coincide with those of the principal axes of inertia \citep{2013CQGra..30d5009B}.

A rigorous treatment of a n-body multipolar expansion should take into account the different orientation of its axis of symmetry. However, this contribution decrease so quickly that at any accuracy currently attainable it turns out to be an unnecessary complication, since just one planet at a time would give a detectable effect.      

The derivatives of the metric coefficients, with retarded time approximation, have the following expressions:
\begin{eqnarray}
\fl
 {\bar h}_{(a),k} = \left(- \frac{n_{(a)}^k}{r_{(a)}^2}(1+ 2 n_{(a)} \cdot  \tilde v) +\frac{\tilde v^k}{r^2_{(a)}} \right)  \left[ 1-\sum_{m=2}^{\infty}  J_m \left( \frac{R_{(a)}}{r_{(a)}} \right)^m P_m(\cos \theta_{(a)}) \right]   \nonumber \\
\fl
+\frac{1}{r_{(a)}} (1+  n_{(a)} \cdot  \tilde v)\left \{ \sum_{m=2}^{\infty}  J_m  \left( \frac{ R_{(a)}}{r_{(a)}} \right)^m  \left[ m (1+  n_{(a)} \cdot  \tilde v)^{m} P_m(\cos \theta_{(a)})  \right. \right. \nonumber \\
\fl
\left. \left. 
\left( \frac{n^k_{(a)}}{r_{(a)}}   
-   \frac{\tilde v^k}{r_{(a)}}  (1+ \delta_{ij} n^i_{(a)}  \tilde v^j)^{-1} \right)- P_m(\cos \theta_{(a)})_{,k} \right] \right \} \label{eq:h_par_i},
\end{eqnarray}
\begin{eqnarray}
\fl
{\bar h}_{(a),0} = \left( \frac{n_{(a)}  \cdot \tilde {\beta}}{r^2_{(a)}}   \right) (1+ n_{(a)} \cdot \tilde {\beta}_{(a)})   \left[ 1- \sum_{m=2}^{\infty}  J_m \left( 
\frac{ R_{(a)}}{r_{(a)}} \right)^m  P_m (\cos \theta_{(a)}) \right]  \nonumber \\
\fl
+ \frac{1}{r_{(a)}} (1+ n_{(a)} \cdot  \tilde v) \left[- \sum_{m=2}^{\infty}  J_m   \left( \frac{ R_{(a)}}{ r_{(a)}} \right)^m  \left(  P_m(\cos \theta_{(a)})_{,0} \right. \right . \nonumber \\
\fl +\left. \left. m   P_m(\cos \theta_{(a)} ) \frac{n_{(a)}  \cdot \tilde {\beta}}{r^2_{(a)}}  (1+ n_{(a)} \cdot  \tilde v)^m\right) \right] ,
\label{eq:h_par_0}
\end{eqnarray}
where $n^i=r^i/r$ . 
A general n-body solution should include the multipolar structure of the sources.  Nevertheless, according to the current astrometric accuracy and for an oblate body,  the quadrupole approximation can be considered enough \citep[see][]{2003AJ....125.1580K}. If we omit the higher multipole moments and restrict ourselves only to $m=2$,  denoting by $s^k$  the axis of the sources which is normal to the source equatorial plane, $P_m(\cos \theta_{(a)}) $ is approximated as:
\begin{eqnarray}
\fl
P_2(\cos \theta_{(a)})=  \frac{3 (s_{(a)} \cdot r(\sigma, \tilde \sigma'))^2 }{2 r(\sigma, \tilde \sigma')^2} -\frac{1}{2} 
\approx  \frac{3}{2}  (s_{(a)} \cdot  n_{(a)} - s_{(a)} \cdot \tilde v)^2 (1+ n_{(a)} \cdot \tilde v)^2-\frac{1}{2}, 
\end{eqnarray}
which to first order in $\tilde v^i$ becomes
 \begin{equation}
 \fl
P_2(\cos \theta_{(a)}) =  \frac{3}{2}   (s_{(a)} \cdot  n_{(a)})^2 (1+ 2 n_{(a)} \cdot \tilde v)-  3  (s_{(a)} \cdot  n_{(a)}) ( s_{(a)} \cdot \tilde v) -\frac{1}{2} + O(\tilde v^2); 
\end{equation}
therefore
\begin{eqnarray}
 \fl
 {\bar h}_{(a),k} =- \frac{n_{(a)}^k}{r_{(a)}^2}(1+ 2 n_{(a)} \cdot  \tilde v) +\frac{\tilde v^k}{r^2_{(a)}}  +   J_2  R_{(a)}^2  \left\{  \frac{3n_{(a)}^k}{r_{(a)}^4}   \left[ \frac{5 (s_{(a)}   \cdot  n_{(a)})^2}{2} +9 (s_{(a)}   \cdot  n_{(a)})^2  \right. \right. \nonumber \\ 
 \fl
 (n_{(a)} \cdot  \tilde v)   -\frac{1}{2} - \left.  \left.3 (s_{(a)} \cdot  n_{(a)})(s_{(a)} \cdot  \tilde v)- \frac{2}{3}(n_{(a)} \cdot  \tilde v) \right]  \right.\nonumber \\
\fl \left.+ \frac{\tilde v^k}{2r_{(a)}^4}  \left[ 1 - 3(s_{(a)}   \cdot  n_{(a)})^2 \right] +\frac{3s_{(a)}^k}{r_{(a)}^4} [  (s_{(a)} \cdot  \tilde v) -  (s_{(a)}   \cdot  n_{(a)})  (1 + 3n_{(a)} \cdot  \tilde v )] \right\} \nonumber\\
 \fl +  O(\tilde v^2),
\label{eq:h_par_q_i}
\end{eqnarray}
and
\begin{eqnarray}
\fl
 {\bar h}_{(a),0} &=&
 \frac{n_{(a)}  \cdot {\tilde v}}{r^2_{(a)}}    +  \frac{3J_2  R_{(a)}^2 }{ r_{(a)}^4} \left[  \frac{n_{(a)}  \cdot \tilde v}{2}  -  \frac{5 }{2} (n_{(a)}  \cdot \tilde v)  (s_{(a)}   \cdot  n_{(a)})^2 \right] +  O(\tilde v^2).
\label{eq:h_par_q_0}
\end{eqnarray}

By taking into account the target accuracy of Gaia \citep[see][]{2003AJ....125.1580K,2008PhRvD..77d4029L}, the velocity contributions for an oblate body should be neglected. However,  for sake of consistency and completeness with the assumptions adopted in this work, neglecting, a priori, terms which are part of the solution is not justified, even if the application to Gaia will surely dismiss many of them. Probably a Gaia-like mission that achieves few sub-microarsecond in accuracy will benefit of these analytical contributions, especially in regards of a cross-checking comparison between different approaches. In this case it would be better to consider a metric which properly contemplate all the complexities  of a non-spherical gravitational body; that, at the moment, is out of the scope of the present paper and deserves a dedicated work (see, e.g., \cite{2013CQGra..30d5009B}).  

Therefore, the RAMOD reduced master equations which take into account the quadrupole structure for the $a$-th single source finally become:

\begin{itemize}

\item RAMOD3aQ (R3aQ) 
\begin{eqnarray}
\fl
\frac{\mathrm{d}\bar{\ell}^{k}_{(a)}}{\mathrm{d}\sigma} = (\mathrm{R3a})^k +   \frac{2 J_2  R_{(a) }^2 \mathcal{M}_{(a)}}{r_{(a)}^4}  \left\{ 3 \left[d^{k}_{(a)} - \frac{ \bar \ell^k (\bar \ell \cdot n_{a})} {2}\right]  \left[\frac{5 (s_{(a)}   \cdot  n_{(a)})^2}{2}  \right.  \right. \nonumber \\
\fl + 9 (s_{(a)}   \cdot  n_{(a)})^2 (n_{(a)} \cdot  \tilde v)   - \frac{1}{2} - \left. \left. 3 (s_{(a)}   \cdot  n_{(a)}) (s_{(a)} \cdot  \tilde v) - \frac{2}{3}(n_{(a)} \cdot  \tilde v) \right]   \right. \nonumber \\
\fl + \frac{1}{2} \left[  d^{k}_{v} - \frac{\bar \ell^k (\bar \ell \cdot \tilde v)}{2}\right] \left[1- 3 (s_{(a)}   \cdot  n_{(a)})^2   \right] + \left. 3 \left[ d^{k}_{s}  -\frac{\bar \ell^k (\bar \ell \cdot s_{a})} {2} \right] \left[ (s_{(a)} \cdot  \tilde v)  \right. \right. \nonumber \\
\fl \left. \left. -  (s_{(a)}   \cdot  n_{(a)})  (1+ 3 (n_{(a)} \cdot  \tilde v )) \right]\right\} + O(\tilde v^2)  +  O(h^2) \equiv  (\mathrm{R3aQ})^k,
\label{eq:meq-q-r3a}
\end{eqnarray}

where $d^k_{s_{(a)}}= s^k_{(a)} - \bar{\ell}^{k} ( \bar\ell \cdot s_{(a)}) $. 

In case of velocity equal to zero, RAMOD equations become:    
\begin{eqnarray}
\fl 
\frac{\mathrm{d}\bar{\ell}^{k}_{(a)}}{\mathrm{d}\sigma} -(\mathrm{R3s})^k  =  \frac{2 J_2  R_{(a) }^2 \mathcal{M}_{(a)}}{r_{(a)}^4}  \left\{   \frac{3}{2} \left[ d^{k}_{(a)} -\frac{ \bar \ell^k (\bar \ell \cdot n_{a})} {2} \right]   \left[5 (s_{(a)}   \cdot  n_{(a)})^2- 1 \right]  \right. \nonumber \\
\fl  \left. +3   \left[ \frac{\bar \ell^k (\bar \ell \cdot s_{a})} {2}- d^{k}_{s} \right]  (s_{(a)}   \cdot  n_{(a)}) \right\}+  O(h^2)
\equiv  (\mathrm{R3sQ})^k \label{eq:meq-q-r3af-zerov}
\end{eqnarray}
\item RAMOD3bQ (R3bQ)
\begin{eqnarray}
\fl 
\frac{d\bar\ell^0_{(a)}}{d\sigma}= (\mathrm{R3b})^0  + \frac {3 \mathcal{M}_{(a)}J_2  R_{(a)}^2 }{ 2r_{(a)}^4} \left[  \frac{n_{(a)}  \cdot \tilde v}{2}  -  \frac{5 }{2} (n_{(a)}  \cdot \tilde v)  (s_{(a)}   \cdot  n_{(a)})^2\right]  \nonumber \\
\fl + O(\tilde v^2)+  O(h^2) \equiv (\mathrm{R3bQ})^0 \label{eq:meq0-q-r3bf} \\
&{}&\nonumber\\
\fl \frac{d\bar\ell^k_{(a)}}{d\sigma} =(\mathrm{R3aQ})^k -  \bar\ell^k  (\mathrm{R3bQ})^0+  O(\tilde v^2)+  O(h^2)  \equiv (\mathrm{R3bQ})^k
\label{eq:meqk-q-r3bf}
\end{eqnarray}
\item RAMOD4aQ  (R4aQ) 
\begin{eqnarray}
\fl
\frac{d\bar\ell^0_{(a)}}{d\sigma} = (\mathrm{R3bQ})^0  -\frac {4 \mathcal{M}_{(a)}J_2  \tilde R_{(a)}^2 } {r^4_{(a)}}  (\bar \ell \cdot \tilde v) \left\{  \frac{3}{2} (\bar \ell \cdot n_{a})  \left[5 (s_{(a)}   \cdot  n_{(a)})^2 -1 \right]  \right.  \nonumber \\
\fl \left.  - 3 (\bar \ell \cdot s_{a})(s_{(a)}   \cdot  n_{(a)}) \right\} +  O(\tilde v^2)+  O(h^2)\equiv (\mathrm{R4aQ})^0
\label{eq:meq0-q-r4af} \\
&{}&\nonumber\\
\fl \frac{d\bar\ell^k_{(a)}}{d\sigma}= (\mathrm{R3bQ})^k - \frac {4 \mathcal{M}_{(a)}J_2  \tilde R_{(a)}^2 } {r^4_{(a)}}  \left\{\frac{3}{2} (\bar \ell \times n_{a} \times \tilde v )^k  \left[5(s_{(a)}   \cdot  n_{(a)})^2 -1 \right]  \right.   \nonumber\\
\fl  \left. +  3 (\bar \ell \times \tilde v  \times s_{a})^k (s_{(a)}   \cdot  n_{(a)}) \right\} +  O(\tilde v^2)+  O(h^2)\equiv (\mathrm{R4aQ})^k.
 \label{eq:meqk-q-r4af}
\end{eqnarray}
\end{itemize}

\section{Light propagation through the Solar System and parametrized trajectories}
\label{sec4}

When a photon approaches the weak gravitational field of the Solar System, heading to a Gaia-like observer, it will feel the gravitational field generated by the mass of the bodies of the system while it will be rather insensitive to the contribution to the field due to their own motion. If one compares the scale of the Solar System and the photon crossing time through it - approximately 10 hours in total- the gravitational field of the Solar System cannot significantly change in a dynamical sense during such time, to the point that the source velocity can be considered constant all along the photon trajectory. This last remark facilitates the solution of the RAMOD equations.

Let us make explicit the vorticity of the congruence $\mathbf{u}$:
\begin{eqnarray}
\omega_{\rho \sigma}&=& P^{\alpha}_{\rho}(u) P^{\beta}_{\sigma}(u) \nabla_{[\alpha} u_{\beta]} \nonumber \\
&=& \nabla_{[\rho} u_{\sigma]} + u_{[\rho} \dot{u}_{\sigma]}
\label{eq:vorticity}
\end{eqnarray}
where $ P^{\alpha}_{\rho}(u)= \delta^{\alpha}_{\rho} + u^{\alpha} u_{\rho}$ is the metric induced on each hypersurfaces of simultaneity of $\mathbf{u}$. Considering that $u^{\alpha}  u_{\alpha}=-1$, and $u^{\alpha} \nabla_{\alpha} u^{\beta}= \dot{u}^{\beta}$, we deduce:
\begin{eqnarray}
\omega_{\rho \sigma}&=& -\eta_{0 [\rho} \partial_{\sigma]} h_{00} + \partial_{[\rho} h_{\sigma]0} + \partial_0 (\eta_{0 [\rho}  h_{\sigma]0})\nonumber 
\end{eqnarray}
which implies
\begin{eqnarray}
\omega_{00}&=& 0\nonumber \\
\omega_{0i}&=& 0\nonumber \\
\omega_{ij}&=&\partial_{[i} h_{j]0}.
\label{eq:vor-spa}
\end{eqnarray}
Taking into account the metric (\ref{eq:me-simplify}), equations (\ref{eq:vor-spa}) show that if we want a vanishing vorticity we have to choose $ h \vec{\nabla} \times \vec{\tilde v} + \vec{\nabla}(h) \times  \vec{\tilde v} = 0$, which is satisfied if  the velocity of the source is zero, i.e. a static case, or is constant, which corresponds to the case remarked above about the variations of the gravitational fields of the Solar System as regards the photon crossing time through it.

Now, within the scale of a vorticity-free geometry,  from the Frobenius theorem, the space-time can be foliated and one can always map the whole geodesic onto the hypersurface of simultaneity of the local barycentric observer at the time of observation.
In this case the mapped trajectory can be expressed in a parametrized form with respect to the centre-of-mass (CM) of the gravitationally bounded system \citep{2011CQGra..28w5013C}: 
\begin{equation}
x^i= \hat{\xi^i}+ \int_o^{\hat\tau} \bar{\ell}^i d\hat{\tau},
\label{eq:parametrization}
\end {equation}
where
\begin{itemize}
\item $\hat{\xi}^i$ is the impact parameter with respect to the centre-of-mass of the gravitationally bounded system, {\it i.e.}, $ \delta_{ij} \bar{\ell}^{i} \hat{\xi}^j = 0 +  \mathcal{O}(h)$ at the point of the closest approach with modulus $\hat \xi = \delta_{ij} {\hat \xi}^{i} \hat{\xi}^j$;   
\item  $\hat \tau = \sigma - \hat \sigma$, being $\hat \sigma$ the value of geodesic parameter at the point of the closest approach.
\end{itemize}

Furthermore, if we approximate the quantity $\bar{\ell}^i$ in terms of small perturbations with respect to the unperturbed light direction $\bar \ell_{\not {0}}$:
 \begin{equation}
 \bar{\ell}^i = \bar{\ell}^{i}_{\not 0} + \delta \bar{\ell}^{i} + (\delta \bar{\ell}^{i})^2 + ...
\label{eq:lo-exp}
\end {equation}
we note that the term $(\delta \bar{\ell}^{i}) $ can be neglected in the master equations (\ref{eq:me0}) and (\ref{eq:mek}) because of the order of $O(h)$ (see also \citep{2006CQGra..23.5467D}).
This implies that equation (\ref{eq:parametrization}), for our purpose, can be approximated as:
\begin{equation}
x^i= \hat{\xi^i}+  (\bar{\ell}^i_{\not 0} +\delta \bar\ell^i) \hat{\tau} + O\left(h^2 \right) ,
\label{eq:parametrization-2}
\end {equation}
where $\delta \bar\ell^j$  is of the order of the deflection, {\it i.e.} $\epsilon^2$.

\begin{figure*} \centering
\includegraphics[width=15cm]{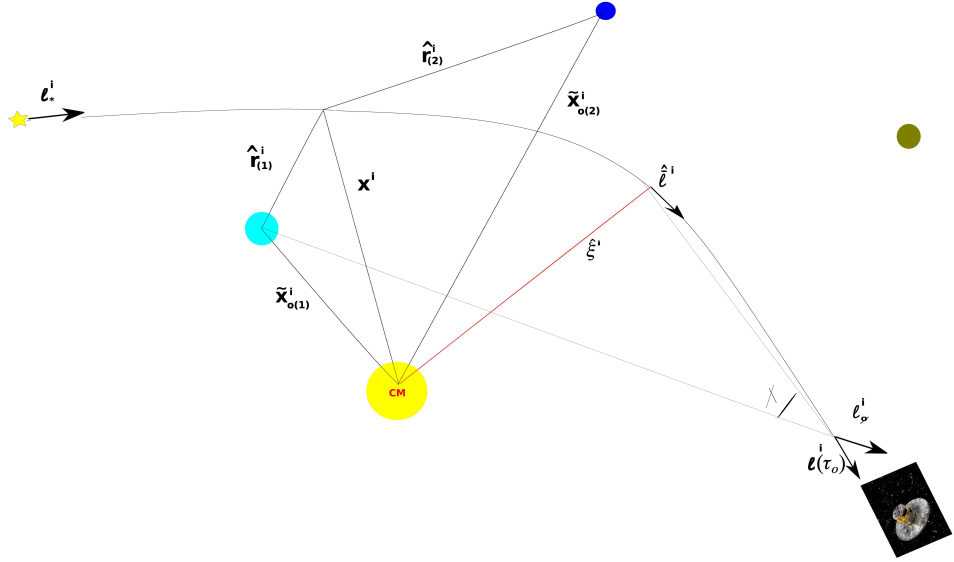} \caption{Mapped trajectories and relative source positions $\hat r^i_{(a)} =x^i - \tilde x^i_{(a)} $ on the slice at the time of observation $(\tau_o)$; $\hat{\xi}^i$ is the impact parameter with respect to the centre-of-mass CM of the gravitationally bounded system,  $\bar{\ell}_{\not 0}$ the unperturbed light direction, whereas $\bar{\ell}_{*}$ that one at the star, and $\chi$ the angle between the directions at the observer towards the point of maximum approach and the centre-of-mass of the source.} \label{fig:may}
\end{figure*}

Finally, after these assumptions, the approximated retarded distance can be parametrized as 
\begin{equation}
\mathbf{r}^i(\sigma, \tilde \sigma')  \approx \hat{\xi^i}+  (\bar{\ell}^i_{\not 0} +\delta \bar\ell^i) \hat{\tau}  - \tilde x^i(\tilde \sigma)  - \tilde v^i (\tilde \sigma) r (\sigma, \tilde \sigma). 
\label{eq:final_ret_dis}
\end{equation}
and the distance $r^{i}_{(a)} $ can be reformulated as
\begin{equation}
 r^i(\sigma, \tilde \sigma') \approx  r^i (\hat \tau) - \tilde v(\tilde \sigma)^i  \hat r,
\end{equation}
 where
\begin{equation}
 r^i (\hat \tau) =\hat{\xi}^i+ (\bar{\ell}^i_{\not 0}+ \delta \bar\ell^k ) \hat\tau  - \tilde{x}^i(\tilde \sigma)=   \hat r_p + (\bar{\ell}^i_{\not 0}+ \delta \bar\ell^k )  \hat\tau ,
\label{eq:rjo}
\end{equation}
namely the relative distance on the slice at the time of observation without the contribution of the source velocity, while 
\begin{equation}
 \hat r_p^i = \hat{\xi}^i - \tilde{x}^i,
\end{equation}
is the relative distance between the point of maximum approach of the photon to the CM  and the source centre-of-mass.   
In figure (\ref{fig:may}) it is sketched the relationship among these vectors on the slice corresponding to the time of observation. 

By using the scalar ($\cdot$) and vectorial ($\times$) products and the parametrization,  the impact parameter (\ref{eq:sip-g}) with respect to the source becomes:
\begin{equation}
 d(\hat \tau)^k= [\bar \ell \times (r(\hat \tau) \times \bar \ell)]^k .
\label{eq:planet-impact-parameter}
\end{equation}
Note that it does not depends on $\hat \tau$ by definition, so it coincides with 
\begin{equation}
 d^k_p= [\bar \ell \times (\hat r_p \times \bar \ell)]^k
\label{eq:planet-impact-parameter}
\end{equation}
or, with the approximation (\ref{eq:lo-exp}), 
 \begin{equation}
 d^k_p= \hat r_p^k-  \bar \ell^k_{\not 0} (\hat r_p \cdot  \bar \ell_{\not 0}),
\label{eq:planet-impact-parameter-st}
\end{equation}
with modulus $d_p = \sqrt{ d_p \cdot d_p}$.
Moreover,  we can assume that $\hat \tau/d_p \approx1/  \tan \chi$, where $\chi$ is the angle between the directions at the observer towards the point of maximum approach and the centre-of-mass of the source.
In this  case it is
\begin{equation}
 r (\hat \tau)\approx \frac{d_p}{\sin \chi} \sqrt{1+  \frac{(\bar{\ell}_{\not 0} \cdot \hat r_p )^2 }{ d_p^2} \sin \chi^2 +2 \frac{(\bar{\ell}_{\not 0} \cdot \hat r_p )}{ d_p} \cos \chi \sin \chi}.
\label{eq: approxrchi}
\end {equation}    

\section{Light deflection by spherical and oblate spheroid gravitational sources with constant velocity}
\label{sec5}
Condition (\ref{eq:vor-spa}) constrains the solution of the geodesic equation to specific circumstances. In particular we do not consider equations $\mathrm{R4a^k}$ and $\mathrm{R4aQ^k}$ since they derive from the terms $\partial_{[i} h_{j]0}$, which are null because of our physical assumption that the sources move with constant velocity.   
As far as the light deflection is concerned, we expect that the velocity contributions become relevant in affecting light propagation in the case of close approach when general relativistic effects become of the order of Gaia's expected accuracy together with the multipolar structure of the source.
In this section we proceed to make the solutions for light deflection explicit according to the RAMOD classification.

 \subsection{Monopole contribution without velocity: R3s case}
This solution includes only the spatial derivative of the metric as in the static case.

Then equation (\ref{eq:meq-r3s}) can be integrated as follows:

\begin{eqnarray}
\fl
\mathrm{\Delta}\bar{\ell}^{k}_{R3s}&=& 2  \sum_{a} \mathcal{M}_{(a)} 
\left \{ 
\int_{\hat \tau}^{\hat \tau_o}   \left [\frac{1}{2} \bar{\ell}^{k}_{\not 0}  (\bar{\ell}_{\not 0} \cdot {r}) - d^k_p  \right]  \frac{d \hat \tau }{r^3}  \right \}+ O \left( h^2 \right),
\label{eq:red2-master-equation-qbodya-monopole}
\end{eqnarray}
namely:
\begin{eqnarray}
\fl 
\mathrm{\Delta}\bar{\ell}^{k}_{R3s}&=&2\sum_{a} \mathcal{M} _{(a)}
\left \{ 
  \left [\frac{1}{2} \bar{\ell}^{k}_{\not 0}  (\bar{\ell}_{\not 0} \cdot \hat r_{p}) - 
d_p^k  \right]  \int_{\hat \tau}^{\hat \tau_o}  \frac{d \hat \tau }{r^3}  +  \frac{1}{2} \bar{\ell}^{k}_{\not 0} \int_{\hat \tau}^{\hat \tau_o}  \frac{ \hat \tau d \hat \tau }{\hat r^3}  \right \} + O \left( h^2 \right),
\label{eq:red2-master-equation-qbodya-m}
\end{eqnarray}
i.e. (see appendix A for the list of the solved integrals)
\begin{eqnarray}
\fl \mathrm{\Delta}\bar{\ell}^{k}_{R3s}&=& 2 \sum_{a} \mathcal{M}_{(a)} \frac{1}{d_p^2} 
\left \{ 
  \left [ \frac{1}{2} \bar{\ell_{\not 0}}^{k}  ({\bar{\ell_{\not 0}}} \cdot  {\hat r}_p) - d_p^k   \right]   \left[  ({\bar\ell}_{\not 0} \cdot  { n})   \right]^{\hat \tau_o}_{\hat \tau}-
  - \frac{1}{2} \bar{\ell_{\not 0}}^{k} \left[ ({ n} \cdot   {\hat r}_{p}) \right]^{\hat \tau_o}_{\hat \tau} \right \}
\label{eq:master-equation-static-monopole}
\end{eqnarray}
which reduces to 
\begin{eqnarray}
\fl \mathrm{\Delta}\bar{\ell}^{k}_{R3s}&=&2 \sum_{a} \mathcal{M}_{(a)} 
\left \{ - \frac{\bar{\ell_{\not 0}}^{k} }{2}  \left[  \frac{1}{  r}\right]^{\hat \tau_o}_{\hat \tau}-  \frac{d_p^k}{d_p^2}  \left[ {\bar \ell}_{\not 0} \cdot  { n}  \right]^{\hat \tau_o}_{\hat \tau}  \right \} + O \left( h^2 \right).
\label{eq:red9-master-equation-n-body}
\end{eqnarray}

Formula (\ref{eq:red9-master-equation-n-body}) can be easily converted in the one found by Klioner (2003). If one considers a source at infinity ($\bar \ell_{\not 0}^k \equiv  \sigma^k$):

\begin{eqnarray}
\Delta \bar \ell^k_{R3s} &\approx & 2 \sum_{a} \mathcal{M} 
\left \{ -  \frac{\sigma^{k} }{ 2 r(\hat \tau_o) } - \frac{d_p^k}{ d_p^2}  \left[ 1  + {\sigma}\cdot  { n}(\hat \tau_o) \right]  \right \}+ O \left( h^2 \right).
\label{eq:klo}
\end{eqnarray}
Then, if the observable shift w.r.t. to the direction at infinity is  $ \delta \sigma^k = c^{-1} [\sigma \times \Delta \dot x \times \sigma ]^k$, we get
\begin{eqnarray}
\delta \sigma^k &\approx &2 \sum_{a} \mathcal{M} 
\left \{ -  \frac{d_p^k}{ d_p^2}  \left[ 1  +   {\sigma}\cdot  { n}(\hat \tau_o)  \right] 
 \right \} +O \left( h^2 \right).
\label{eq:klo-def}
\end{eqnarray}

\subsubsection{Monopole light deflection due to one body.}

Let us apply  equation (\ref{eq:meq-r3s}) to the computation of the light deflection due to one body, for example, the Sun, and check if the well known formula is recovered.

With respect to the local barycentric observer $u^{\alpha} = e^{\phi} \delta^{\alpha}_0$, the total light deflection is given by the modulus:
\begin{equation}
\Delta \bar \ell = \sqrt {P(u)_{\alpha \beta}  \Delta \bar \ell^{\alpha}   \Delta \bar \ell^{\beta}}= \sqrt {(g_{\alpha \beta} + u_{\alpha} u_{\beta}) \Delta \bar \ell^{\alpha}   \Delta \bar \ell^{\beta}}.
\label{eq:total-deflection}
\end{equation}
We have to consider only the euclidean metric $ \delta_{ij} $, since $\Delta \bar \ell^k$ is  of the order of $h$, moreover it is already projected with respect to $u^{\alpha}$, and in the static case $ \bar \ell^{0}=0$; therefore, we limit the integration to the following expression
 \begin{equation}
 \Delta \bar{\ell}^k = 2  \mathcal{M}^{(sun)} \left[ \frac{1}{2}  \int_{-\infty}^{\infty}\frac{\bar{\ell}^{k}_{o} \hat \tau } {(\hat{\xi}^2 + \hat{\tau}^2)^{3/2}}  \mathrm{d}\hat \tau-  \int_{-\infty}^{\infty} \frac{\hat{\xi}^k }{(\hat{\xi}^2 + \hat{\tau}^2)^{3/2}}\mathrm{d}\hat \tau \right], \label{eq:total-deflection-sun}
\end{equation} 
which  yields the well known solution:
\begin{equation}
\Delta \bar{\ell} =  \frac{4 }{ \hat{\xi}} \mathcal{M}^{(sun)}. 
\end{equation}
We can also check the validity of our assumption by limiting the integration at the observer. Considering the source at infinity, from (\ref{eq: approxrchi}) 
\begin{eqnarray}
\fl \mathrm{\Delta}\bar{\ell}^{k}_{R3s}&=&2 \sum_{a} \mathcal{M}_{(a)} 
\left \{ - \frac{\bar{\ell_{\not 0}}^{k} }{2} \left[ \frac{\sin \chi}{d_p \sqrt{1+  \frac{(\bar{\ell}_{\not 0} \cdot \hat r_p )^2 }{d_p^2} \sin \chi^2 +2 \frac{(\bar{\ell}_{\not 0} \cdot \hat r_p )}{d_p} \cos \chi \sin \chi} }\right] \right. \nonumber \\
\fl &-& \left. \frac{d_p^k}{d_p^2}  \left[   \frac{({\bar \ell}_{\not 0} \cdot  {\hat r_p} + \hat \tau_o) \sin \chi}{d_p \sqrt{1+  \frac{(\bar{\ell}_{\not 0} \cdot \hat r_p )^2 }{d_p^2} \sin \chi^2 +2 \frac{(\bar{\ell}_{\not 0} \cdot \hat r_p )}{d_p} \cos \chi \sin \chi} }  + 1 \right] \right \} +O \left( h^2 \right),
\end{eqnarray}
and in the limit $\chi << 1$, since $ \hat \tau_o/d_p =\cos \chi/\sin \chi$, we obtain 
\begin{eqnarray}
\Delta \bar{\ell}^k  &\approx & 2 \sum_{a}  \mathcal{M}_{(a)} \left \{ - 2 \frac{d_p^k}{ d_p^2}  
 \right \} +O \left( h^2 \right).
\label{eq:onedef}
\end{eqnarray}

Then the modulus in the case of one body is again:
\begin{equation}
\Delta \bar{\ell} =  \frac{4}{\hat{\xi}} \mathcal{M}. 
\end{equation}

These results check the validity of RAMOD master equation when it is applied to one single body and make us confident to proceed with the solution for a n-body system.

\subsection{Monopole and velocities contribution: R3a case}

This solution includes the perturbations depending on the spatial derivative of the $h$ term and the retarted distance approximation. According to the assumption of section \ref{sec4} we consider also the velocity of the source constant, i.e. not dependent on $\hat \tau$. For sake of simplicity, let us denote:
\begin{equation}
D^k_p= d^{k}_{p}- \frac{\bar \ell^k_{\not 0} (\bar \ell_{\not0} \cdot \hat r_{p})} {2}
\end{equation}
\begin{equation}
D^k_v=d^{k}_{v} -\frac{\bar \ell^k_{\not 0} (\bar \ell_{\not0} \cdot v)} {2}. 
\end{equation}

In the case discussed here we limit ourself to the integration of the following terms:
\begin{eqnarray}
\fl \mathrm{\Delta}\bar{\ell}^{k}_{R3a}&\approx &  \Delta \bar \ell^k_{R3s}+  2  \sum_{a} \mathcal{M}_{(a)}  \left \{ 
 - 2 D_p^k  (\hat{r}_p \cdot \tilde v)  \int_{\hat \tau}^{\hat \tau_o}  \frac{d \hat \tau }{r^4} +  \left[- 2 D_p^k  (\bar \ell_{\not 0} \cdot \tilde v) + \bar \ell_{\not 0}^k  (\hat{r}_p \cdot \tilde v)  \right]  \right. \nonumber \\
\fl && \left.  \int_{\hat \tau}^{\hat \tau_o} \frac{\hat \tau d \hat \tau }{ r^4}+ \bar \ell_{\not 0}^k  (\bar \ell_{\not 0} \cdot \tilde v)  \int_{\hat \tau}^{\hat \tau_o} \frac{\hat \tau^2 d \hat \tau }{r^4} + D_v^k    \int_{\hat \tau}^{\hat \tau_o}  \frac{d \hat \tau }{ r^2} \right \}.
\label{eq:meq-r3a-k-ret}
\end{eqnarray}

In addition to the static case, the solution of the R3a master equation is:
\begin{eqnarray}
\fl \mathrm{\Delta}\bar{\ell}^{k}_{R3a}& = &  \Delta \bar \ell^k_{R3s}+  2 \sum_{a}  \frac{ \mathcal{M}_{(a)}}{d^2_p}   \left \{ 
 -  d^2_p \frac{\bar \ell^k_{\not 0}}{2} (\bar \ell_{\not 0} \cdot \tilde v)  \left[ \frac{\hat \tau }{r^2}  \right]^{\hat \tau_o}_{\hat \tau} -   D^k_p (\hat{r}_p \cdot \tilde v)  \left[ \frac{ (\bar \ell_{\not 0} \cdot  n)}{ r}  \right]^{\hat \tau_o}_{\hat \tau}  \right. \nonumber \\  
 \fl &+& \left.   \left[  D_v^k d_p -  \frac{D^k_p } { d_p}  ({d}_p \cdot \tilde v)+ \frac{\bar \ell^k_{\not 0}}{2d_p } \left((\tilde v \times \hat r_p) \cdot (\hat r_p \times \bar \ell_{\not 0})\right) \right] \left[\arctan \left(\frac{\bar \ell_{\not 0} \cdot  r} { d_p}\right) \right]^{\hat \tau_o}_{\hat \tau}  \right. \nonumber \\ 
\fl  &+&  \left.  \left[  D_p^k   (\bar \ell_{\not 0} \cdot \tilde v) - \frac{\bar \ell^k_{\not 0}}{2}  ({d}_p \cdot \tilde v)  \right]  \left[ \frac{ (\hat{r}_p \cdot  n)}{ r}  \right]^{\hat \tau_o}_{\hat \tau}\right \}+  O(h) + O(v^2)
\label{eq:r3a-sol}
\end{eqnarray}

If the source is at infinity and in the limit of $\chi << 1$, i.e. at the observer and for grazing rays,  these contributions vanish (see appendix A). 

\subsection{Monopole and velocities contribution: R3b case}
This solution includes the perturbations depending on the time derivative of the $h$ term according to the retarted distance approximation. Assuming  the source velocity constant, we have to integrate: 
 \begin{eqnarray}
\Delta \bar\ell^0_{R3b} &\approx &     \sum_{a} \mathcal{M}_{(a)}  \left \{   ( \hat r_p \cdot \tilde v)   \int_{\hat \tau}^{\hat \tau_o}  \frac{d \hat \tau }{ r^3}  +  ( \bar \ell_{\not 0} \cdot \tilde v)  \int_{\hat \tau}^{\hat \tau_o}  \frac{\hat \tau d \hat \tau }{ r^3}  \right \}\label{eq:meq0-r3bf}\\
&{}&\nonumber\\
 \Delta \bar \ell^k_{R3b} &\approx &  \Delta \bar \ell^k_{R3a}  -  \bar\ell^k_{\not 0}   \Delta \bar \ell^0_{R3b},
 \label{eq:meqk-r3bf}
\end{eqnarray}
The solution is straightforwardly (appendix A):
\begin{eqnarray}
\Delta \bar\ell^0_{R3b} &\approx &     \sum_{a}  \frac{\mathcal{M}_{(a)}}{d_p^2}  \left \{  ( \hat r_p \cdot \tilde v)   \left[  (\bar \ell_{\not 0} \cdot n)\right]^{\hat \tau_o}_{\hat \tau}  -  ( \bar \ell_{\not 0} \cdot \tilde v)  \left[  (n \cdot \hat r_p) \right]^{\hat \tau_o}_{\hat \tau}  \right \}\label{eq:meq0-r3b-sol1}\\
&{}&\nonumber\\
 \Delta \bar \ell^k_{R3b} &\approx &  \Delta \bar \ell^k_{R3a}  -  \bar\ell^k_{\not 0}   \Delta \bar \ell^0_{R3b}.
 \label{eq:meqk-r3b-sol1}
\end{eqnarray}
For the source at infinity,  in the limit of the small angle $\chi$ between the directions at the observer towards the source center of mass and the point of closest approach, from appendix A these contributions result: 
\begin{eqnarray}
\Delta \bar\ell^0_{R3b} &\approx &    2  \sum_{a} \mathcal{M}_{(a)}    \frac{( \hat r_p  \cdot  \tilde v)} {d_p^2}  \label{eq:meq0-r3b-sol}\\
&{}&\nonumber\\
 \Delta \bar \ell^k_{R3b} &\approx &  4  \sum_{a} \mathcal{M}_{(a)} 
\left \{ \frac{d_p^k}{ d_p^2}  \left[ 1  +  \frac{ (\bar \ell_{\not 0} \cdot \hat n_p)}  { 2}   \right] 
   -  \bar\ell^k_{\not 0}      \frac{( \hat r_p  \cdot  \tilde v)} {2 d_p^2}   \right \}  .
 \label{eq:meqk-r3b-sol}
\end{eqnarray}

\subsection{Monopole and velocities contribution: R4a case}
The $\mathrm{R4a}^k$ equations include the perturbations depending on the spatial derivative of the $h_{0i}$ term, i.e. those depending on the mass-current contribution. According to the constrain on the vorticity imposed by the physical assumption as discussed in section (\ref{sec4}), we have to consider only equation $\mathrm{R4a}^0$: 

\begin{eqnarray}
\fl \Delta \bar\ell^0_{R4a}& \approx&  \Delta \bar\ell^0_{R3b} + 4  \sum_{a}  \mathcal{M}_{(a)} ( \bar\ell_{\not 0} \cdot \tilde{v})  \left[ ( \bar\ell_{\not 0} \cdot \hat r_p)   \int_{\hat \tau}^{\hat \tau_o}\frac{ d \hat \tau }{ r^3} + \int_{\hat \tau}^{\hat \tau_o}\frac{\hat \tau d \hat \tau }{\hat r^3}   \right]  \label{eq:meq0-r4afc}\\
\fl \Delta \bar \ell^k_{R4a}&\approx & \Delta \bar \ell^k_{R3b}
\label{eq:meqk-r4-afc}
\end{eqnarray}
i.e.

\begin{eqnarray}
\fl \Delta \bar\ell^0_{R4a}& \approx&  \Delta \bar\ell^0_{R3b} + 4  \sum_{a}   \frac{\mathcal{M}_{(a)}}{d_p^2}( \bar\ell_{\not 0} \cdot \tilde{v})  \left\{ ( \bar\ell_{\not 0}  \cdot \hat r_p) \left[ (\bar \ell_{\not 0} \cdot n)  \right]^{\hat \tau_o}_{\hat \tau}- \left[  ( n \cdot \hat r_p) \right]^{\hat \tau_o}_{\hat \tau}   \right\}  \label{eq:meq0-r4-sol}\\
\fl \Delta \bar \ell^k_{R4a}&\approx & \Delta \bar \ell^k_{R3b}
\label{eq:meqk-r4-sol}
\end{eqnarray}

Again for a source placed at infinity and in the limit of a grazing light ray, we get at the observer:
\begin{eqnarray}
\Delta \bar\ell^0_{R4a}& \approx& 2  \sum_{a} \frac{\mathcal{M}_{(a)} }{d_p^2}  \left \{ ( \hat r_p  \cdot  \tilde v)   + 4  ( \bar\ell_{\not 0} \cdot \tilde{v})  ( \bar\ell_{\not 0}  \cdot \hat r_p) \right\} \label{eq:meq0-r4-sol-infi}\\
&{}&\nonumber\\
\Delta \bar \ell^k_{R4a}&\approx & 4  \sum_{a} \frac{\mathcal{M}_{(a)} } { d_p^2} 
\left \{ d_p^k \left[ 1  +  \frac{ (\bar \ell_{\not 0} \cdot \hat n_p)}  { 2}   \right] 
   -  \bar\ell^k_{\not 0}      \frac{( \hat r_p  \cdot  \tilde v)} {2 }   \right \}.
\label{eq:meqk-r4-sol-infi}
\end{eqnarray}

\subsection{Light deflection by a static oblate body: R3sQ case}
For sake of convenience, in this  we drop the suffix $(a)$. Considering that the quadrupole structure of the source may be relevant in affecting light propagation only during close approach, here and after, to compute the quadrupole contribution, we refer only to a single body.  
Moreover, as in section (5.2), we denote:
\begin{equation}
D^k_s=d^{k}_{s} -\frac{\bar \ell^k_{\not 0} (\bar \ell_{\not0} \cdot s)} {2}, 
\end{equation}
where $s^k$ is the axis of the source normal to its equatorial plane.
Equation (\ref{eq:meq-q-r3af-zerov}) in function of the parametrized trajectory can be integrated as follows
\begin{eqnarray}
\fl \Delta \bar \ell^k_{R3sQ} &=&   3 J_2  R^2 \mathcal{M} \left\{ -\left[ D^k_p + 2 (s \cdot \hat r_p)  D^{k}_{s} \right] \int_{ \hat \tau}^{\hat \tau_o}\frac{ d \hat \tau }{ r^5 } +   5(s \cdot \hat r_p)^2 D^{k}_{p}  \int_{\hat  \tau}^{ \hat \tau_o}\frac{ d \hat  \tau }{\hat r^7 }\right. \label{eq:meq-q-r3as-zero-v} \\
\fl &+&\left. \left[  \bar \ell^k_{\not0} - 2 (s \cdot \bar \ell_{\not0})D^{k}_{s} \right] \int_{ \hat \tau}^{\hat  \tau_o}\frac{\hat  \tau d\hat  \tau }{ r^5 }+ 5 (s\cdot \hat r_p) \left[ 2 (s\cdot \bar \ell_{\not0}) D^{k}_{p} -   \bar \ell^k_{\not0} (s\cdot \hat r_p)  \right ] \int_{\hat \tau}^{\hat  \tau_o}\frac{ \hat \tau d \hat  \tau }{  r^7 } \right. \nonumber \\
\fl &+&\left. 5 (s\cdot \bar \ell_{\not0}) \left[(s\cdot \bar \ell_{\not0}) D^{k}_{p} - \bar \ell^k_{\not0}(s\cdot \hat r_p) \right] \int_{\hat \tau}^{\hat  \tau_o}\frac{ \hat \tau^2 d \hat  \tau }{ r^7 }-  5 \bar \ell^k_{\not0} (s\cdot \bar \ell_{\not0})^2 \int_{ \hat \tau}^{\hat \tau_o}\frac{\hat  \tau^3 d\hat  \tau }{ r^7 } 
 \right\} + O(h^2),\nonumber
\end{eqnarray}

\begin{table}[htp]
\caption{  \label{tab:listintegral} List of the recurrent integrals }
   \begin{indented}
    \item[]\begin{tabular}{@{} l l} 
    \br
       \bf{Notation}   &  \bf{Corresponding expressions }   \\
      \mr
      $ r(\hat \tau) $  &  $ \sqrt{\hat r_p^2 +2 (\bar \ell_{\not 0} \cdot   \hat r_p) \hat \tau+ \hat \tau^2 } $ \\
       $I_{0/0} $          & $\arctan[(\bar \ell_{\not 0} \cdot   r)/ d_p]$  \\
       $ I_{1/0}$  &  $  \mathrm{Log}(r \cdot \bar\ell_{\not 0}+  r) $ \\
       $I_{0/1}$           &$ 1/r$     \\
       $I_{0/2}$     &  $1/r^2$ \\
      $I_{0/3}$     &  $1/ r^3$ \\
      $I_{0/4}$     &  $1/ r^4$ \\
      $I_{0/5}$     &  $1/ r^5$ \\
      $I_{0/5}$     &  $1/r^6$ \\
      $I_{1/1}$      & $ (\bar \ell_{\not 0} \cdot  r)/   r $ \\
      $I_{1/2}$      & $ (\bar \ell_{\not 0} \cdot  r)/   r^2 $   \\
      $I_{1/3}$      &  $ (\bar \ell_{\not 0} \cdot   r)/  r^3 $ \\
      $I_{1/4}$      &   $ (\bar \ell_{\not 0} \cdot  r)/ r^4 $\\
      $I_{1/5}$      &   $ (\bar \ell_{\not 0} \cdot  r)/  r^5 $\\
      $I_{1/6}$      &   $ (\bar \ell_{\not 0} \cdot   r)/  r^6 $\\
      $I_{2/1}$      &   $ ( r \cdot  \hat r_p)/   r $\\
      $I_{2/2}$      &   $ ( r \cdot  \hat r_p)/  r^2 $\\
      $I_{2/3}$      &   $ ( r \cdot  \hat r_p)/  r^3 $\\
      $I_{2/4}$      &   $  ( r \cdot  \hat r_p)/   r^4$  \\
      $I_{2/5}$  &   $  ( r \cdot  \hat r_p)/  r^5$  \\
      $I_{2/6}$  &   $  ( r \cdot  \hat r_p)/   r^6$  \\
         \br
   \end{tabular}
   \end{indented}
\end{table}

namely (for the details of the solution see table \ref{tab:listintegral} and  Appendix A),
\begin{eqnarray}
\fl \Delta \bar\ell^k_{R3sQ}&=&   6 J_2  R^2 \frac {\mathcal{M}}{ d_p^2} \left\{\left[I_{0/3}\right]_{\hat \tau}^{\hat \tau_o} C_1^k + \left[\hat \tau I_{0/3}\right]_{\hat \tau}^{\hat \tau_o} C_2^k  + \left[\hat  \tau I_{0/5}\right]_{\hat \tau}^{\hat \tau_o} C_3^k +  \left[I_{1/1}\right]_{\hat \tau}^{\hat \tau_o}  C_4^k \right. \nonumber \\
\fl &+ &\left.   \left[I_{1/3}\right]_{\hat \tau}^{\hat \tau_o} C_5^k  +  \left[I_{1/5}\right]_{\hat \tau}^{\hat \tau_o} C_6^k +  \left[I_{2/3}\right]_{\hat \tau}^{\hat \tau_o} + C_7^k\left[I_{2/5}\right]_{\hat \tau}^{\hat \tau_o} C_8^k \right\} + O(h^2),
\label{eq:meq-q-r3as-zero-v-brutta}
\end{eqnarray}
If we consider a grazing light ray emitted by a star located at infinity, above equations reduce to:
\begin{equation}
\Delta \bar\ell^k_{R3sQ}=6   J_2  R^2 \frac{\mathcal{M}}{d_p^2} \left\{ 2 C_4^k \right\} + O(h^2).\label{eq:meq-q-r3as-zero-v-grazing}
\end{equation}

Similarly to the spherical case, let us compute the total light deflection close to one gravity source.   
In this case, we expect that our formula reduces to the available expressions known in the literature \citep{PhysRevD.22.2947, 2006CQGra..23.4853C, 2008PhRvD..77d4029L, 2013CQGra..30d5009B}.
For this scope we can assume:
\begin{itemize}
\item  $ d_p^k = \hat r_p^k \equiv \xi^k  $  the radial vector;
\item  $ d_p^k/d_p = - n^k  $  the unit radial vector;
\item  $ \bar \ell_{\not o}^k \equiv t^k$   the tangential vector to the line-of-sight;
\item $ \sqrt{\xi^i \xi_i} \equiv  d_p $ the impact parameter,
\item and $m^k= (\bar \ell \times n)^k$  the orthoradial direction.
\end{itemize}
Then:
\begin{eqnarray}
\fl \Delta \bar\ell^k_{R3sQ}&=&  4   J_2  R^2 \frac{\mathcal{M}}{d_p^3} \left\{ n^k [1-(s\cdot t)^2 - 4 (s\cdot n)^2 ]  +  2 (s\cdot n) [  s^k - t^k(s\cdot t)] \right\} + O(h^2),
\label{eq:meq-q-r3as-zero-v-grazing}
\end{eqnarray}
and, by expressing the vector $s^k$ as linear combination in terms of the orthonormal basis ($\mathbf{t, n,m}$)  
\begin{equation}
s^k= (s \cdot t) t^k +  (s \cdot n) n^k + (s \cdot m) m^k, 
\end{equation}
we obtain the same formula deduced in Crosta and Mignard \citep{2006CQGra..23.4853C}:
\begin{eqnarray}
\fl \Delta \bar \ell^k &=& \Delta \bar\ell^k_{R3s} + \Delta \bar\ell^k_{R3sQ} \nonumber \\
\fl &=& \frac{4 \mathcal{M}}{d_p}\left\{ \left[ 1 +   \frac{J_2  R^2}{d_p^2} \left( 1-(s\cdot t)^2 - 2 (s\cdot n)^2 \right)\right] n^k +  \frac{ J_2  R^2}{d_p^2}   (s \cdot m) (s \cdot n)m^k \right\}
\label{eq:meq-q-r3as-zero-v-grazing}
\end{eqnarray}

\subsection{Quadrupole contribution with velocity: R3aQ case}

Equation (\ref{eq:meq-q-r3a}) in function of the parametrized trajectory becomes:
\begin{eqnarray}
\fl \Delta \bar\ell^k_{R3aQ}= \Delta \bar\ell^k_{R3a} +   \Delta \bar\ell^k_{R3sQ} \label{eq:meq-q-r3qv-integral} \\
\fl  +  2 J_2  R^2 \mathcal{M} \left\{ \left[ \frac{1}{2} D^{k}_{v} +3 D^k_s (s \cdot \tilde v) \right] \int_{\hat \tau}^{\hat \tau_o}\frac{ d \hat \tau }{r^4 } - \left[D^{k}_{p} \left( 9 (s \cdot \hat r_p) (s \cdot \tilde v) + 2(\hat r_p \cdot \tilde v)\right)  \right. \right. \nonumber\\
\fl- \frac{3}{2}D^{k}_{v}  (s \cdot \hat r_p)^2   + \left.\left. 3 D^{k}_{s} (s \cdot \hat r_p) (\hat r_p \cdot \tilde v)  \right]  \int_{\hat \tau}^{\hat \tau_o}\frac{ d \hat \tau }{ r^6 } + 27 D^{k}_{p} (s \cdot \hat r_p)^2 (\hat r_p \cdot \tilde v) \int_{\hat \tau}^{\hat \tau_o}\frac{ d \hat \tau }{ r^8 }  \right. \nonumber \\
\fl \left. - \left[D^{k}_{p} \left( 9 (s\cdot \bar \ell_{\not0})  (s \cdot \tilde v) +2 (\bar \ell_{\not0} \cdot \tilde v)\right) -  \bar \ell^k_{\not 0} \left( 9 (s\cdot \hat r_p)  (s \cdot \tilde v)+2 (\hat r_p \cdot \tilde v) \right) + 3D^{k}_{v}  (s\cdot \bar \ell_{\not0})(s\cdot \hat r_p)  \right. \right. \nonumber  \\
\fl \left. \left.   +3 D^{k}_{s}  \left( (\hat r_p \cdot \tilde v) (s\cdot \bar \ell_{\not0})  + (s\cdot \hat r_p)  (\bar \ell_{\not0} \cdot \tilde v)  \right)\right] \int_{\hat \tau}^{\hat \tau_o}\frac{\hat \tau d \hat \tau }{ r^6 }+ 27 (s\cdot \hat r_p) \left[ D^{k}_{p} \left((s\cdot \hat r_p)(\bar \ell_{\not0} \cdot \tilde v)\right. \right. \right. \nonumber \\
\fl \left.\left. \left. +2  (s\cdot \bar \ell_{\not0})  (\hat r_p \cdot \tilde v)\right)  -  \bar \ell^k_{\not0}  (s\cdot  \hat r_p) (\hat r_p \cdot \tilde v)\right] \int_{\hat \tau}^{\hat \tau_o}\frac{\hat \tau d \hat \tau }{ r^8 } +\left[ 9 \bar \ell^k_{\not0} (s\cdot \bar \ell_{\not0}) (s\cdot \tilde v) + 2\bar \ell^k_{\not0}(\bar \ell_{\not0}\cdot \tilde v)   \right.\right.  \nonumber \\
\fl \left.  \left.  -3 D^{k}_{s} (\bar \ell_{\not0}\cdot \tilde v) \left( (s\cdot \bar \ell_{\not0})  + (s\cdot \hat r_p)  \right) + \frac{1}{2} D_v^k (s\cdot \bar \ell_{\not 0})^2 \right] \int_{\hat \tau}^{\hat \tau_o}\frac{\hat \tau^2 d \hat \tau }{r^6 } + 27 \left[(s\cdot \bar \ell_{\not0}) D^{k}_{p}   \right. \right. \nonumber\\
\fl \left. \left.  + \left(2 (s\cdot \hat r_p) (\bar \ell_{\not0} \cdot \tilde v)+ (s\cdot \bar \ell_{\not0}) (\hat r_p \cdot \tilde v) \right)  -   \bar \ell^k_{\not0} (s\cdot \hat r_p) + \left( (s\cdot \hat r_p)(\bar \ell_{\not0} \cdot \tilde v)\right. \right.\right. \nonumber\\ 
\fl \left. \left. \left. +2 (s\cdot \bar \ell_{\not0})(\hat r_p \cdot \tilde v)\right)\right] \int_{\hat \tau}^{\hat \tau_o}\frac{\hat \tau^2 d \hat \tau }{r^8 } +  27 (s\cdot \bar \ell_{\not0})\left[ D^{k}_{p} (s\cdot \bar \ell_{\not0}) (\bar \ell_{\not0} \cdot \tilde v)   -   \bar \ell^k_{\not0} \left( 2(s\cdot \hat r_p)  (\bar \ell_{\not0}\cdot \tilde v)  \right. \right. \right. \nonumber \\
\fl \left.\left. \left.  + (s\cdot \bar \ell_{\not0}) (\hat r_p \cdot \tilde v)\right)\right] \int_{\hat \tau}^{\hat \tau_o}\frac{\hat \tau^3 d \hat \tau }{ r^8 } + 27 \left[  \bar \ell^k_{\not0} (s\cdot \bar \ell_{\not0})^2 (\bar \ell_{\not0} \cdot \tilde v) \right] \int_{\hat \tau}^{\hat \tau_o}\frac{\hat \tau^4 d \hat \tau }{ r^8 } \right\} + O(\tilde v^2)+ O(h^2). \nonumber
\end{eqnarray}
%where the solution of the integrals can be found in appendix A.
Namely, from table 1 and appendix A, we denote
\begin{eqnarray}
\fl \Delta \bar\ell^k_{R3aQ}= \Delta \bar\ell^k_{R3a} +   \Delta \bar\ell^k_{R3sQ}+   J_2  R^2 \frac {\mathcal{M}}{ d_p^2} 
\left\{
  \left[I_{0/0}\right]_{\hat \tau}^{\hat \tau_o}  C_9^k + \left[I_{0/4}\right]_{\hat \tau}^{\hat \tau_o} C_{10}^k +   \left[\hat \tau I_{0/4}\right]_{\hat \tau}^{\hat \tau_o} C_{11}^k   \right. \nonumber \\
\fl \left.+ \left[I_{0/6}\right]_{\hat \tau}^{\hat \tau_o} C_{12}^k   + \left[\tau I_{0/6}\right]_{\hat \tau}^{\hat \tau_o} C_{13}^k   + \left[I_{1/2}\right]_{\hat \tau}^{\hat \tau_o}  C_{14}^k + \left[I_{1/4}\right]_{\hat \tau}^{\hat \tau_o} C_{15}^k +  \left[I_{1/6}\right]_{\hat \tau}^{\hat \tau_o} C_{16}^k\right. \nonumber \\
\fl \left.+  \left[ I_{2/4}\right]_{\hat \tau}^{\hat \tau_o} C_{17}^k +\left[I_{2/6}\right]_{\hat \tau}^{\hat \tau_o} C_{18}^k \right\} + O(h^2) + O(\tilde v^2). \label{eq:meq-q-r3as-zero-v-coeff}
\end{eqnarray}
For a source at infinity and in the limit of $\chi <<1$ the above contributions vanish as in the case of mass monopole moving with constant velocities.  

\subsection{Quadrupole contribution with velocity: R3bQ case}

From equations (\ref{eq:meqk-q-r3bf}) we have to solve

\begin{eqnarray}
\fl \Delta \bar\ell^0_{R3bQ} =  \Delta \bar\ell^0_{R3b}  + 3 \mathcal{M}J_2  R^2 \left\{
\frac{\hat r_p \cdot \tilde v}{2} \int_{\hat \tau}^{\hat \tau_o}\frac{ d \hat \tau }{r^5 } + \frac{\bar \ell_{\not 0} \cdot v}{2} \int_{\hat \tau}^{\hat \tau_o}\frac{\hat \tau d \hat \tau }{ r^5 } \right. \label{eq:meq-q-r3ab-velocity}  \\
\fl \left.- \frac{5}{2} \left[  (\hat r_p \cdot \tilde v) (s \cdot \hat  r_p)^2 \right] \int_{\hat \tau}^{\hat \tau_o}\frac{ d \hat \tau }{ r^7 }  - 5 (s\cdot \hat r_p) \left[  (\hat r_p \cdot \tilde v) (s\cdot \bar \ell_{\not0}) +\frac{1}{2} (\bar \ell_{\not 0} \cdot \tilde v) (s\cdot \hat r_p)  \right ] \int_{\hat \tau}^{\hat \tau_o}\frac{\hat \tau d \hat \tau }{r^7 }  \right. \nonumber \\
\fl \left. - 5(s\cdot \bar \ell_{\not0}) \left[  \frac{1}{2}(s\cdot \bar \ell_{\not0}) (\hat r_p \cdot \tilde v) +  (\bar \ell_{\not 0} \cdot \tilde v)(s \cdot \hat  r_p ) \right] \int_{\hat \tau}^{\hat \tau_o}\frac{\hat \tau^2 d \hat \tau }{ r^7 } \right. \nonumber \\
\fl \left. -\frac{5}{2}  \left[(s\cdot \bar \ell_{\not0})^2  (\bar \ell_{\not 0} \cdot \tilde v) \right]  \int_{\hat \tau}^{\hat \tau_o}\frac{\hat \tau^3 d \hat \tau }{ r^7 }\right\}+ O(\tilde v^2)+  O(h^2)\nonumber \\
\fl \Delta \bar\ell^k_{R3bQ} =\Delta \bar\ell^k_{R3aQ}-  \bar\ell^k_{\not 0}  (\mathrm{R3bQ})^0+  O(\tilde v^2)+  O(h^2).
\label{eq:meqk-q-r3bf-velocity}
\end{eqnarray}
Expressing the above formulae in terms of the solutions of the integrals listed in table 1, these quantities read
\begin{eqnarray}
\fl \Delta \bar\ell^0_{R3bQ} = \Delta \bar\ell^0_{R3b}  + \frac{ \mathcal{M}J_2  R^2}{d^2_p} \left\{
\left[ I_{0/3}\right]_{\hat \tau}^{\hat \tau_o} C_{19} +\left[ \hat \tau I_{0/3}\right]_{\hat \tau}^{\hat \tau_o} C_{20}  + \left[ \hat \tau I_{0/5}\right]_{\hat \tau}^{\hat \tau_o} C_{21} + \left[ I_{1/1}\right]_{\hat \tau}^{\hat \tau_o}  \right. \nonumber \\
\fl \left.C_{22}  +\left[ I_{1/3}\right]_{\hat \tau}^{\hat \tau_o} C_{23} +  \left[ I_{1/5}\right]_{\hat \tau}^{\hat \tau_o} C_{24}+ \left[ I_{2/3}\right]_{\hat \tau}^{\hat \tau_o} C_{25} +\left[ I_{2/5}\right]_{\hat \tau}^{\hat \tau_o} C_{26}
\right\}+O(\tilde v^2)+  O(h^2)
\label{eq:meq-q-r3ab-velocity}\\
\fl \Delta \bar\ell^k_{R3bQ} =\Delta \bar\ell^k_{R3aQ} -  \bar\ell^k_{\not 0}  \Delta \bar\ell^0_{R3bQ}+  O(\tilde v^2)+  O(h^2).
\label{eq:meqk-q-r3bf-velocity}
\end{eqnarray}

For a stellar source at infinity and in the limit of light ray grazing the gravity source, we get:

\begin{eqnarray}
\Delta \bar\ell^0_{R3bQ} &=&\Delta \bar\ell^0_{R3b}  +\frac{2 \mathcal{M}J_2  R^2}{d^2_p}  C_{22}
 +  O(\tilde v^2)+  O(h^2)
\label{eq:meq-q-r3ab-velocity}\\
&{}&\nonumber\\
\Delta \bar\ell^k_{R3bQ} &=&\Delta \bar\ell^k_{R3aQ} -  \bar\ell^k_{\not 0} \Delta \bar\ell^0_{R3bQ}+  O(\tilde v^2)+  O(h^2).
\label{eq:meqk-q-r3bf-velocity}
\end{eqnarray}
With the same notations of the single static source, the above equations collapse to 
\begin{eqnarray}
\fl \Delta \bar\ell^0_{R3bQ} =  \Delta \bar\ell^0_{R3b}  +  \frac{\mathcal{M}J_2  R^2 }{d_p^3}\left\{ 
 - (n \cdot \tilde v) [ 1-4  (s\cdot n)^2 - (s \cdot t)^2]  +  2 (s \cdot t) (s \cdot n) ( t\cdot v)  \right\} \nonumber \\ +  O(\tilde v^2)+  O(h^2)
\label{eq:meq-q-r3ab-velocity-velocity}\\
\fl \Delta \bar\ell^k_{R3bQ} =\Delta \bar\ell^k_{R3aQ} +   \frac{\mathcal{M}J_2  R^2 }{d_p^3}\left\{ t^k  (n \cdot \tilde v) [ 1-4  (s\cdot n)^2 - (s \cdot t)^2]  -   2 t^k (s \cdot t) (s \cdot n) ( t\cdot v) \right\} \nonumber \\
+ O(\tilde v^2)+  O(h^2)
\label{eq:meqk-q-r3bf-velocity-velocity-cm}
\end{eqnarray}

\subsection{Quadrupole contribution with velocity: R4aQ solution}
The integral expressions of equations (\ref{eq:meq0-q-r4af}) and (\ref{eq:meqk-q-r4af}) with the physical assumption of a source moving with constant velocity are:

\begin{eqnarray}
\fl \Delta \bar\ell^0_{R4aQ} =  \Delta \bar\ell^0_{R3bQ}  -4  \mathcal{M}_{(a)}J_2   R_{(a)}^2 (\bar \ell_{\not0} \cdot \tilde v)  \left\{
- \left[ \frac{1}{2} (\bar \ell_{\not 0} \cdot \hat r_p) +  3 (s\cdot \bar \ell_{\not0}) (s \cdot \hat r_p) \right] \int_{\hat \tau}^{\hat \tau_o} \frac{ d \hat \tau }{ r^5 } \right. \nonumber \\
\fl - \left.  \left[   \frac{1}{2} + 3 (s\cdot \bar \ell_{\not0})^2 \right] \int_{\hat \tau}^{\hat \tau_o}\frac{\hat \tau d \hat \tau }{ r^5 }+  \left[\frac{5}{2} (s \cdot \hat r_p)^2 (\bar \ell_{\not 0} \cdot \hat r_p)  \right] \int_{\hat \tau}^{\hat \tau_o}\frac{ d \hat \tau }{ r^7 }  + \frac{5}{2} (s\cdot \hat r_p) \left[  2 (s\cdot \bar \ell_{\not0})\right. \right. \nonumber \\
\fl  \left. \left. (\bar \ell_{\not 0} \cdot \hat r_p) +   (s \cdot \hat r_p) \right ] \int_{\hat \tau}^{\hat \tau_o}\frac{\hat \tau d \hat \tau }{ r^7 }+  \frac{5}{2}(s\cdot \bar \ell_{\not0}) \left[ (\bar \ell_{\not0} \cdot \hat r_p)(s\cdot \bar \ell_{\not0}) + 2  (s \cdot \hat r_p ) \right] \int_{\hat \tau}^{\hat \tau_o}\frac{\hat \tau^2 d \hat \tau }{ r^7 }\right. \nonumber \\
\fl + \left. \frac{5}{2}  (s\cdot \bar \ell_{\not0})^2  \int_{\hat \tau}^{\hat \tau_o}\frac{\hat \tau^3 d \hat \tau }{ r^7 }\right\} +  O(\tilde v^2)+  O(h^2)
\label{eq:meq0-q-r4af-velocity}\\
\fl \Delta \bar\ell^k_{R3bQ}=\Delta \bar\ell^k_{R4aQ} +  O(\tilde v^2)+  O(h^2)
 \label{eq:meqk-q-r4-velocity}
\end{eqnarray}
Considering table 1 and appendix A, the solution reads:
\begin{eqnarray}
\fl \Delta \bar\ell^0_{R4aQ} =  \Delta \bar\ell^0_{R3bQ}  - \frac{2  \mathcal{M}_{(a)}J_2   R_{(a)}^2}{d^2_p} (\bar \ell_{\not0} \cdot \tilde v)  \left\{
\left[ I_{0/3}\right]_{\hat \tau}^{\hat \tau_o} C_{27}+  \left[ \hat \tau I_{0/3}\right]_{\hat \tau}^{\hat \tau_o} C_{28} +
\left[ \hat \tau I_{0/5}\right]_{\hat \tau}^{\hat \tau_o}  \right. \nonumber \\
\fl \left.C_{29}  + \left[ I_{1/1}\right]_{\hat \tau}^{\hat \tau_o} C_{30} +
\left[ I_{1/3}\right]_{\hat \tau}^{\hat \tau_o} C_{31} + \left[ I_{1/5}\right]_{\hat \tau}^{\hat \tau_o} C_{32}+ \left[ I_{2/3}\right]_{\hat \tau}^{\hat \tau_o} C_{33} + \left[ I_{2/5}\right]_{\hat \tau}^{\hat \tau_o} C_{34}
\right\} \nonumber \\
+  O(\tilde v^2)+  O(h^2)
\label{eq:meq0-q-r4af-velocity-int}\\
&{}&\nonumber\\
\fl \Delta \bar\ell^k_{R4aQ}=\Delta \bar\ell^k_{R3bQ} 
+  O(\tilde v^2)+  O(h^2)
 \label{eq:meqk-q-r4-velocity-int}
\end{eqnarray}

For a source located at infinity and in the limit of the small angle $\chi$ between the directions at the observer towards the body center-of-mass and the point of closest approach of the light trajectory, we obtain :
\begin{eqnarray}
\fl \Delta \bar\ell^0_{R4aQ} =  \Delta \bar\ell^0_{R3bQ}  - \frac{4  \mathcal{M}_{(a)}J_2   R_{(a)}^2}{d^2_p} (\bar \ell_{\not0} \cdot \tilde v)   C_{30} +   O(\tilde v^2)+  O(h^2)
\label{eq:meq0-q-r4af-velocity-int} \\
\fl \Delta \bar\ell^k_{R4aQ}= \Delta \bar\ell^k_{R4aQ} +  O(\tilde v^2)+  O(h^2).
 \label{eq:meqk-q-r4-velocity-int}
\end{eqnarray}
Again, with the same conventions of the static source, for the time component we deduce finally for the light deflection:
\begin{eqnarray}
\fl \Delta \bar\ell^0_{R4aQ} = \Delta \bar\ell^0_{R3bQ} + \frac{16  \mathcal{M}_{(a)}J_2   R_{(a)}^2}{d^3_p} (t  \cdot \tilde v) (s\cdot n)  (s \cdot t) +   O(\tilde v^2)+  O(h^2)
\label{eq:meq0-q-r4af-velocity-int-inf-st}.
\end{eqnarray}

\section{Photon trajectory}
\label{sec6}

The first integration of RAMOD equations gives the estimates of the deflection effects. The second integration solves the ray tracing for the photon emitted by a star and intercepted at the observer's location, thus obtaining the quantity $ \mathrm{\Delta} x^{\alpha}= x^{\alpha}(\hat \tau_o) - x^{\alpha}(\hat \tau_*)$ at the different approximation levels. In general since the first member is $dx^{\alpha}/d \hat \tau - \bar \ell_{o}^k$, the integral to be solved in the case of monopoles with constant velocities became:

\begin{eqnarray}
\fl \mathrm{\Delta} x^{k}_{R3s} =\bar \ell^k_{o} \Delta \hat \tau + 2 \sum_{a} \mathcal{M}_{(a)} 
\left \{ - \frac{\bar{\ell_{\not 0}}^{k} }{2}\int^{\hat \tau_o}_{\hat \tau_*}  I_{0/1} d \hat \tau 
 -  \frac{d_p^k }{d_p^2} \int^{\hat \tau_o}_{\hat \tau_*}I_{1/1}    d \hat \tau  \right \} 
+ O \left( h^2 \right),
\label{eq:red9-master-equation-n-body-traj}
\end{eqnarray}

\begin{eqnarray}
\fl \mathrm{\Delta}x^{k}_{R3a}=   \Delta x^k_{R3s}+ 2  \sum_{a} \frac{\mathcal{M}_{(a)}}{d_p^2}   \left \{ -  d^2_p \frac{\bar \ell^k_{\not 0}}{2} (\bar \ell_{\not 0} \cdot \tilde v)  \int^{\hat \tau_o}_{\hat \tau_*} \hat \tau  I_{0/2} d \hat \tau  -  \frac{D^k_p}{2}  ( \hat r_p \cdot \tilde v) \int^{\hat \tau_o}_{\hat \tau_*}  I_{1/2} d \hat \tau \right.   \nonumber \\
 \fl  \left. + \left[  D_v^k d_p -  \frac{D^k_p } { d_p}  (d_p \cdot \tilde v) +  \frac{\bar \ell^k_{\not 0}}{2d_p } \left((\tilde v \times \hat r_p) \cdot (\hat r_p \times \bar \ell_{\not 0})\right)
  \right]   \int^{\hat \tau_o}_{\hat \tau_*}   I_{0/0} d \hat \tau  \right.  \nonumber \\
 \fl \left.+ \left[  D_p^k   (\bar \ell_{\not 0} \cdot \tilde v) - \frac{\bar \ell^k_{\not 0}}{2}  (d_p \cdot \tilde v)  \right] \int^{\hat \tau_o}_{\hat \tau_*}   I_{2/2} d \hat \tau\right \}+ O \left( v^2 \right) + O \left( h^2 \right),
\label{eq:r3a-sol-traj}
\end{eqnarray}
\begin{eqnarray}
\fl \Delta x^0_{R3b} =   \bar \ell^0_{o} \Delta \hat \tau  +  \sum_{a} \frac{ \mathcal{M}_{(a)} }{d_p^2} \left \{   ( \hat r_p \cdot \tilde v) \int^{\hat \tau_o}_{\hat \tau_*}I_{1/1}    d \hat \tau  - ( \bar \ell_{\not 0} \cdot \tilde v)\int^{\hat \tau_o}_{\hat \tau_*}  I_{2/1} d \hat \tau \right \} \nonumber \\
 +O \left( v^2 \right)+ O \left( h^2 \right)
\label{eq:meq0-r3b-sol1-traj}\\
\fl  \Delta x^k_{R3b} =  \Delta x^k_{R3a}  -  \bar\ell^k_{\not 0}   \Delta x^0_{R3b}+O \left( v^2 \right) +O \left( h^2 \right),
 \label{eq:meqk-r3b-sol1-traj}
\end{eqnarray}
and
\begin{eqnarray}
\fl \Delta x^0_{R4a} =  \Delta x^0_{R3b} + 4  \sum_{a}  \frac{\mathcal{M}_{(a)}}{d^2_p} ( \bar\ell_{\not 0} \cdot \tilde{v}) \left\{ 
 ( \bar \ell_{\not 0} \cdot \hat r_p) \int^{\hat \tau_o}_{\hat \tau_*}I_{1/1}    d \hat \tau  - \int^{\hat \tau_o}_{\hat \tau_*}  I_{2/1} d \hat \tau \right\} \nonumber \\
+O \left( v^2 \right) +O \left( h^2 \right),
 \label{eq:meq0-r4-soltraj}\\
\fl \Delta x^k_{R4a}=  \Delta x^k_{R3b},
\label{eq:meqk-r4-sol}
\end{eqnarray}
where $\bar \ell^{\alpha}_{o}\equiv\bar \ell^{\alpha} (\hat \tau_o) $ is the line-of-sight direction at the observer used also to solve the boundary value problem \cite{2003CQGra..20.4695B}, and the relationship $\Delta \tau = (\bar \ell_{\not 0} \cdot \Delta r)$ holds.

Then, the trajectory is composed by  the following terms (see appendix A):
\begin{eqnarray}
\fl \mathrm{\Delta} x^{k}_{R3s}=\bar \ell^k _{o} \Delta \hat \tau + 2 \sum_{a} \mathcal{M}_{(a)} 
\left \{ - \frac{\bar{\ell_{\not 0}}^{k} }{2} \left[ \rm{Log}(r \cdot \bar{\ell_{\not 0}} + r ) \right]^{\hat \tau_o}_{\hat \tau_*} -  \frac{d_p^k }{d_p^2} \left[ r \right]^{\hat \tau_o}_{\hat \tau_*}
 \right \} +O \left( h^2 \right),
\label{eq:red9-master-equation-n-body-traj-fin}
\end{eqnarray}

\begin{eqnarray}
\fl \mathrm{\Delta}x^{k}_{R3a}=   \Delta x^k_{R3s}+  2 \sum_{a} \frac{\mathcal{M}_{(a)}}{d^2_p}   \left \{ 
d_p \left[ d_p^k ( \bar \ell_{\not 0} \cdot \tilde v)- \frac{\bar \ell_{\not 0}^k}{2}( \hat r_p \cdot \tilde v) \right]    \left[  \arctan \left(\frac{ \bar \ell_{\not 0} \cdot r}{d_p}\right) \right]^{\hat \tau_o}_{\hat \tau_*}   \right.\nonumber \\
 \fl \left. + \left [d_v^k d_p   -  \frac{D^k_p } { d_p}  (\hat{d}_p \cdot \tilde v) + \frac{\bar \ell^k_{\not 0}}{2d_p } \left((\tilde v \times \hat r_p) \cdot (\hat r_p \times \bar \ell_{\not 0})\right) \right] \left[  ( \bar \ell_{\not 0} \cdot r )  \arctan \left(\frac{ \bar \ell_{\not 0} \cdot  r}{d_p}\right) \right]^{\hat \tau_o}_{\hat \tau_*} \right. \nonumber \\
 \fl \left. - \left[ d^2_p d_v^k + \frac{\bar \ell^k_{\not 0}}{2} \left((\tilde v \times \hat r_p) \cdot (\hat r_p \times \bar \ell_{\not 0})\right) +\frac{\bar \ell^k_{\not 0}}{2} (\hat{d}_p \cdot \tilde v) (\hat r_p \cdot \bar \ell_{\not 0})
 \right]   \left[   \rm{Log}(r)  \right]^{\hat \tau_o}_{\hat \tau_*}\right \}  \nonumber \\
 +O \left( h^2 \right) + O(v^2),
\label{eq:r3a-sol-traj}
\end{eqnarray}

\begin{eqnarray}
\fl \Delta x^0_{R3b} =   \bar \ell^0 _{o} \Delta \hat \tau  +  \sum_{a} \mathcal{M}_{(a)}  \left \{-  ( \bar \ell_{\not 0} \cdot \tilde v)  \left[ \rm{Log}(r \cdot \bar{\ell_{\not 0}} + r ) \right]^{\hat \tau_o}_{\hat \tau_*}  +  \frac{ ( d_p \cdot \tilde v)}{d^2_p} [r]^{\hat \tau_o}_{\hat \tau_*} \right\} \nonumber \\
+O \left( h^2 \right)+ O(v^2)
\label{eq:meq0-r3b-sol1-traj1}\\
\fl \Delta x^k_{R3b} =  \Delta x^k_{R3a}  -  \bar\ell^k_{\not 0}   \Delta x^0_{R3b}+O \left( h^2 \right)+ O(v^2),
 \label{eq:meqk-r3b-sol1-traj2}
\end{eqnarray}
and,
\begin{eqnarray}
\fl \Delta x^0_{R4a} = \Delta x^0_{R3b} - 4  \sum_{a}  \mathcal{M}_{(a)} ( \bar\ell_{\not 0} \cdot \tilde{v}) \left[ \rm{Log}(r \cdot \bar{\ell_{\not 0}} + r ) \right]^{\hat \tau_o}_{\hat \tau_*}  +O \left( h^2 \right) + O(v^2)
\label{eq:meq0-r4-soltraj-boh}\\
\fl \Delta x^k_{R4a}=  \Delta x^k_{R3b} +O \left( h^2 \right) + O(v^2).
\label{eq:meqk-r4-sol}
\end{eqnarray}

Similarly,  the inclusion of the quadrupole terms, at the first order in $h$ and $\tilde v$,  gives additional contributions to the following trajectories (see table 1 and appendix A for the detailed expressions of each term):

\begin{eqnarray}
 \fl \Delta x^k_{R3sQ} =   6 J_2  R^2 \frac {\mathcal{M}}{ d_p^2} 
\left\{
  \left( \frac{C_1^k}{d^2_p} -  \frac{2 C_3^k( \bar\ell_{\not 0} \cdot \hat r_p)}{3d^4_p}+\frac{2 C_8^k}{3d^2_p} \right)  \left[I_{1/1}\right]_{\hat \tau^*}^{\hat \tau_o} -  \frac{C_2^k}{d^2_p}  \left[I_{2/1}\right]_{\hat \tau^*}^{\hat \tau_o} -  \frac{C_3^k}{3d^2_p}   \right. \nonumber \\
 \fl \left.  \left[I_{2/3}\right]_{\hat \tau^*}^{\hat \tau_o}  +  C_4^k [r]_{\hat \tau^*}^{\hat \tau_o}- C_5^k \left[I_{0/1}\right]_{\hat \tau^*}^{\hat \tau_o}  -    \frac{C_6^k}{3} \left[I_{0/3}\right]_{\hat \tau^*}^{\hat \tau_o}
 +  C_7^k \left[ \hat \tau I_{0/1}\right]_{\hat \tau^*}^{\hat \tau_o}  +   \frac{C_8^k}{3}\left[ \hat \tau I_{0/3}\right]_{\hat \tau^*}^{\hat \tau_o} 
 \right\},
\label{eq:meq-q-r3as-zero-v-brutta-traject}
\end{eqnarray}

\begin{eqnarray}
 \fl \Delta x^k_{R3aQ} = \Delta x^k_{R3a}  + \Delta x^k_{R3sQ} +   J_2  R^2 \frac {\mathcal{M}}{ d_p^2} 
\left\{
 \left( ( \bar\ell_{\not 0} \cdot \hat r_p) C_9^k  +   \frac{C_{10}^k}{d^3_p} -  \frac{ C_{11}^k( \bar\ell_{\not 0} \cdot \hat r_p)}{2d^3_p}+\frac{3 C_{12}^k}{8d^5_p} \right. \right. \nonumber \\
\fl \left.\left. - \frac{ 3C_{13}^k( \bar\ell_{\not 0} \cdot \hat r_p)}{8d^5_p} +  \frac{C_{17}^k}{2d_p} + \frac{3 C_{18}^k}{8d^3_p} \right)  \left[I_{0/0}\right]_{\hat \tau^*}^{\hat \tau_o} + C_9^k  \left[ \hat \tau I_{0/0}\right]_{\hat \tau}^{\hat \tau_o} +  \left( C_{14}^k - d_p C_9^k \right)  \right. \nonumber \\
\fl \left.  \left[\mathrm{Log (r)}\right]_{\hat \tau^*}^{\hat \tau_o}+ \frac{1}{d^2_p}  \left(C_{11}^k  + C_{15}^k + ( \bar\ell_{\not 0} \cdot \hat r_p) C_{17}^k \right) \left[I_{2/2}\right]_{\hat \tau^*}^{\hat \tau_o}+  \frac{1}{d^2_p}  \left(C_{10}^k  + \frac{3C_{12}^k}{4d^2_p} - \frac{3 ( \bar\ell_{\not 0} \cdot \hat r_p) C_{13}^k}{4d^2_p}  \right. \right.  \nonumber \\
\fl \left. \left. +  ( \bar\ell_{\not 0} \cdot \hat r_p) C_{15}^k  + r^2_p C_{17}^k + \frac{3C_{18}^k}{4}\right) \left[I_{1/2}\right]_{\hat \tau^*}^{\hat \tau_o}+ \frac{1}{4d^2_p}  \left(C_{12}^k  +  ( \bar\ell_{\not 0} \cdot \hat r_p) C_{16}^k +  r^2_p C_{18}^k  \right) \left[I_{1/4}\right]_{\hat \tau^*}^{\hat \tau_o}  \right. \nonumber \\
\fl \left. - \frac{1}{4d^2_p}  \left(C_{13}^k  +  ( \bar\ell_{\not 0} \cdot \hat r_p) C_{18}^k +  C_{16}^k \right) \left[I_{2/4}\right]_{\hat \tau^*}^{\hat \tau_o} \right\} ,
\label{eq:meq-q-r3as-zero-v-r3qa-traject}
\end{eqnarray}

\begin{eqnarray}
 \fl \Delta x^0_{R3bQ} =   \Delta x^0_{R3b} +   J_2  R^2 \frac {\mathcal{M}}{ d_p^2} 
\left\{  \left( \frac{C_{19}}{d^2_p} - \frac{2 C_{21}( \bar\ell_{\not 0} \cdot \hat r_p)}{3d^4_p} + \frac{2 C_{26}}{3d^2_p} \right)  \left[I_{1/1}\right]_{\hat \tau^*}^{\hat \tau_o}  -  \frac{C_{21}}{d^2_p}  \left[I_{2/1}\right]_{\hat \tau^*}^{\hat \tau_o} \right. \nonumber \\
\fl  - \left.   \left(  \frac{C_{21}}{3d^2_p}  + \frac{2 C_{26}( \bar\ell_{\not 0} \cdot \hat r_p)}{3d^4_p} \right) \left[I_{2/3}\right]_{\hat \tau^*}^{\hat \tau_o}+C_{22} [r]_{\hat \tau^*}^{\hat \tau_o}  - C_{23} \left[I_{0/1}\right]_{\hat \tau^*}^{\hat \tau_o} - \frac{C_{24}}{3} \left[I_{0/3}\right]_{\hat \tau^*}^{\hat \tau_o}  \right. \nonumber \\
\fl  + \left.   C_{25} \left[ \hat \tau I_{0/1}\right]_{\hat \tau^*}^{\hat \tau_o} +\frac{\hat r^2_p C_{26}}{d^2_p}\left[ \hat \tau I_{1/3}\right]_{\hat \tau^*}^{\hat \tau_o}  \right\} ,
\label{eq:meq-q-r3as-zero-v-3bq0-traject} \\
\fl \Delta x^k_{R3bQ} =    \Delta x^k_{R3b} - \bar\ell_{\not 0}^k \Delta x^0_{R3bQ} , 
\label{eq:meq-q-r3as-zero-v-3bqk-traject}
\end{eqnarray}

and, finally, 
\begin{eqnarray}
 \fl \Delta x^0_{R4aQ} =   \Delta x^0_{R3bQ} - 2   J_2  R^2 \frac {\mathcal{M}}{ d_p^2} ( \bar\ell_{\not 0} \cdot \tilde v)
\left\{
  \left( \frac{C_{27}}{d^2_p} -  \frac{2 C_{29}( \bar\ell_{\not 0} \cdot \hat r_p)}{3d^4_p} + \frac{2 C_{34}^k}{3d^2_p} \right)  \left[I_{1/1}\right]_{\hat \tau^*}^{\hat \tau_o} \right. \nonumber\\
 \fl \left. -   \frac{C_{28}}{d^2_p}  \left[I_{2/1}\right]_{\hat \tau^*}^{\hat \tau_o} - \left(  \frac{C_{29}}{3d^2_p}  \right) \left[I_{2/3}\right]_{\hat \tau^*}^{\hat \tau_o}  +C_{30} [r]_{\hat \tau^*}^{\hat \tau_o}  + C_{31}\left[I_{0/1}\right]_{\hat \tau^*}^{\hat \tau_o} - \frac{C_{32}}{3} \left[I_{0/3}\right]_{\hat \tau^*}^{\hat \tau_o}\right. \nonumber \\
 \fl \left.  +  C_{33}^k \left[ \hat \tau I_{0/1}\right]_{\hat \tau^*}^{\hat \tau_o} +  \frac{ C_{34}^k}{3}\left[ \hat \tau I_{0/3}\right]_{\hat \tau^*}^{\hat \tau_o}  \right\},
\label{eq:meq-q-r3as-zero-v-4aq0-traject} \\
\fl \Delta x^k_{R4aQ} =    \Delta x^k_{R3bQ} .  
\label{eq:meq-q-r3as-zero-v-4aq0-traject}
\end{eqnarray}

\section{Conclusions}
The Relativistic Astrometric MODel (RAMOD) is a mathematical tool conceived to model the astrometric measurements  made by an observer in space.
Since its original purpose was to address this problem for the ESA Gaia mission, whose final astrometric accuracy requires the physical model to be accurate at microarcsecond level, RAMOD had to take into account the general relativistic corrections due to the bodies of the Solar System. Despite the apparent straightforwardness of the task and the linearity of the metric given the weak gravitational field regime inside the Solar System, the solution of the inverse ray tracing problem, which allows us to reach the aim above, is rather  intractable unless treated numerically, particularly if retarded time contributions need to be accounted for \citep{2006ApJ...653.1552D}. As far as RAMOD is concerned,  the reason lies mainly on the fact that the main unknown of the differential equations is the observed direction as projected on the rest space of the local barycentric observer and represents {\it locally} what the observer measures of the incoming photons in his/her gravitational environment.
This aspect transforms the geodesic equation into a set of nonlinear coupled differential equations which comprises also 

 for the time component.  
The original version of the RAMOD model was therefore numerical and although successful in its applications with the inclusion of the relativistic satellite attitude \cite{2003CQGra..20.4695B}, it was hard to control and compare with similar astrometric models even with a comprehensive error budget for stellar positions \cite{2008CQGra..25p5015D}. 

Here we present a fully analytical solution of a system of differential equations up to the $\epsilon^3$ level everywhere in the Solar System, which is therefore able to assure a microarcsecond-level accuracy consistent with the precepts of the measurement protocol in General Relativity, and that can also be utilized under observing conditions more demanding than those of the Gaia mission.

The analytical solution is general enough to be applicable to other missions conceived to exploit photon trajectories and extends the analysis of the trajectory perturbations  due to gravitating sources with a non negligible quadrupole structure. 
While the retarded time approximation adopted here and  the solution for the static cases recover the results obtained by similar astrometric models (as proved also in \cite{2011CQGra..28w5013C}), the solutions including the constant velocity of the source give rise to different expressions that deserve to be carefully evaluated in a separated work as were done, for example, in \cite{2010A&A...509A..37C}, and \cite{2014CQGra..31a5021B}. At first glance, in fact, the presence of the time component  $\bar \ell^0$ for the local-line-of-sight has not been contemplated in other models and the $\bar \ell^k$ components do not show  complete coincidence.  A proper comparison, both analytical and numerical,  will precise the physical significance of this unexpected  discrepancy  as one  carries  on with  the   implementation process of the astrometric observables from which the relativistic astrometric parameters  with their appropriate variance and, possibly, covariance values are deduced.

\ack

The authors wish to thank Dr. Donato Bini and Dr. Andrea Geralico for the fruitful discussions and useful comments. This work was supported by the ASI contract I/058/10/0 \&/1.

\bibliographystyle{iopart-num}
\bibliography{mybibl}

\providecommand{\newblock}{}
\begin{thebibliography}{10}
\expandafter\ifx\csname url\endcsname\relax
  \def\url#1{{\tt #1}}\fi
\expandafter\ifx\csname urlprefix\endcsname\relax\def\urlprefix{URL }\fi
\providecommand{\eprint}[2][]{\url{#2}}
% Bibliography created with iopart-num v2.1
% /biblio/bibtex/contrib/iopart-num

\bibitem{2005tdug.conf.....T}
{Turon} C, {O'Flaherty} K~S and {Perryman} M~A~C (eds) 2005 {\em {The
  Three-Dimensional Universe with Gaia}\/}

\bibitem{1987thyg.book.....H}
{Hawking} S~W and {Israel} W (eds) 1987 {\em {Three hundred years of
  gravitation}\/} (Cambridge University Press)

\bibitem{2004ApJ...607..580D}
{de Felice} F, {Crosta} M~T, {Vecchiato} A, {Lattanzi} M~G and {Bucciarelli} B
  2004 {\em Astrophys.\ J.\/} {\bf 607} 580--595

\bibitem{2006ApJ...653.1552D}
{de Felice} F, {Vecchiato} A, {Crosta} M~T, {Bucciarelli} B and {Lattanzi} M~G
  2006 {\em Astrophys.\ J.\/} {\bf 653} 1552--1565 (\textit{Preprint}
  \eprint{arXiv:astro-ph/0609073})

\bibitem{2010ToM.book.....D}
de~Felice F and {Bini} D 2010 {\em {Classical Measurements in Curved
  Space-Times}\/} ({Cambridge University Press})

\bibitem{2010A&A...509A..37C}
{Crosta} M and {Vecchiato} A 2010 {\em Astron. Astrophys.\/} {\bf 509} A37

\bibitem{2003AJ....125.1580K}
{Klioner} S~A 2003 {\em Astron.\ J.\/} {\bf 125} 1580--1597

\bibitem{2012CQGra..29x5010T}
{Teyssandier} P 2012 {\em Classical and Quantum Gravity\/} {\bf 29} 245010
  (\textit{Preprint} \eprint{1206.6309})

\bibitem{2014PhRvD..89f4045H}
{Hees} A, {Bertone} S and {Le Poncin-Lafitte} C {\em Phys.Rev.D\/}

\bibitem{2011CQGra..28w5013C}
{Crosta} M 2011 {\em Classical and Quantum Gravity\/} {\bf 28} 235013
  (\textit{Preprint} \eprint{1012.5226})

\bibitem{1999PhRvD..60l4002K}
{Kopeikin} S~M and {Sch{\"a}fer} G 1999 {\em Phys.\ Rev.\ D\/} {\bf 60} 124002

\bibitem{1990recm.book.....D}
de~Felice F and {Clarke} C~J~S 1990 {\em {Relativity on curved manifolds}\/}
  ({Cambridge University Press})

\bibitem{2013CQGra..30d5009B}
{Bini} D, {Crosta} M, {de Felice} F, {Geralico} A and {Vecchiato} A 2013 {\em
  Classical and Quantum Gravity\/} {\bf 30} 045009

\bibitem{2008PhRvD..77d4029L}
{Le Poncin-Lafitte} C and {Teyssandier} P 2008 {\em Phys.\ Rev.\ D\/} {\bf 77}
  044029 (\textit{Preprint} \eprint{0711.4292})

\bibitem{2006CQGra..23.5467D}
{de Felice} F and {Preti} G 2006 {\em Class.\ Quantum Grav.\/} {\bf 23}
  5467--5476

\bibitem{PhysRevD.22.2947}
Epstein R and Shapiro I~I 1980 {\em Phys. Rev. D\/} {\bf 22}(12) 2947--2949
  \urlprefix\url{http://link.aps.org/doi/10.1103/PhysRevD.22.2947}

\bibitem{2006CQGra..23.4853C}
{Crosta} M~T and {Mignard} F 2006 {\em Class.\ Quantum Grav.\/} {\bf 23}
  4853--4871 (\textit{Preprint} \eprint{arXiv:astro-ph/0512359})

\bibitem{2003CQGra..20.4695B}
{Bini} D, {Crosta} M~T and {de Felice} F 2003 {\em Class.\ Quantum Grav.\/}
  {\bf 20} 4695--4706

\bibitem{2008CQGra..25p5015D}
{de Felice} F and {Preti} G 2008 {\em Classical and Quantum Gravity\/} {\bf 25}
  165015

\bibitem{2014CQGra..31a5021B}
{Bertone} S, {Minazzoli} O, {Crosta} M, {Le Poncin-Lafitte} C, {Vecchiato} A
  and {Angonin} M~C 2014 {\em Classical and Quantum Gravity\/} {\bf 31} 015021
  (\textit{Preprint} \eprint{1306.2367})

\end{thebibliography}

\appendix

\section{Expressions of the coefficients in the quadrupole deflection formulae}

\begin{eqnarray*}
\fl C_1^k=  \frac{ \bar \ell^k_{\not0}(s\cdot \bar \ell_{\not0})^2}{6 d^2_p} \left[ 2 d_p^4- 3  \hat r_p^2 d_p^2 - 3 (\hat r_p \cdot \bar \ell_{\not 0})^4\right]; 
\end{eqnarray*}
%\hline \\
\begin{eqnarray*}
\fl C_2^k =  \frac{\bar \ell^k_{\not0}}{2 d^2_p}(s\cdot \bar \ell_{\not0})^2(\hat r_p \cdot \bar \ell_{\not 0})  \hat r_p^2; 
\end{eqnarray*}
%\hline \\
\begin{eqnarray*}
\fl C_3^k= d^2_p \left\{ - D^{k}_{p}\frac{(s\cdot \bar \ell_{\not0})^2}{2} + \frac{\bar \ell^k_{\not0}}{2}(s\cdot \bar \ell_{\not0}) \left[ (s\cdot \hat r_p)-  2(s\cdot \bar \ell_{\not0})^2 ( \hat r_p \cdot \bar \ell_{\not 0})\right] \right\}; 
\end{eqnarray*}
%\hline \\
\begin{eqnarray*}
\fl C_4^k= \frac{ D^{k}_{p}}{3d_p^2} \left\{ - 1+ 4 \frac{(s\cdot  \hat r_p)^2}{d_p^2  } +  \frac{( \hat r_p \cdot \bar \ell_{\not 0}) (s\cdot \bar \ell_{\not0}) }{d_p^2} \left[ -5  (s\cdot  \hat r_p) - 3( \hat r_p \cdot d_s)\right] + \hat r^2_p \frac{(s\cdot \bar \ell_{\not0})^2}{d_p^2} \right\} \\
\fl + \frac{\bar \ell^k_{\not0}}{3d^2_p}\left\{  -( \hat r_p \cdot \bar \ell_{\not 0})+ \frac{( \hat r_p \cdot \bar \ell_{\not 0})(s \cdot \hat r_p)}{  d_p^2} \left[ (s \cdot \hat r_p)  - 3( \hat r_p \cdot d_s)   \right] + \hat r^2_p \frac{(s\cdot \bar \ell_{\not0})}{d_p^2}  \left[ -(s \cdot \hat r_p)   \right.\right.  \\
\fl \left. \left. + 3(\hat r_p \cdot \bar \ell_{\not 0}) (s\cdot \bar \ell_{\not0})\right]+ \frac{( \hat r_p \cdot \bar \ell_{\not 0})^3(s\cdot \bar \ell_{\not0})^2}{d_p^2} \right\} - \frac{2D^{k}_{s}}{3d_p^2} ( \hat r_p \cdot d_s );
\end{eqnarray*}
%\hline
\begin{eqnarray*}
\fl C_5^k= D^{k}_{p} \left\{- \frac{1}{2} + \frac{2(s\cdot \hat  r_p)^2}{3 d_p^2  }  + \frac{(\hat r_p \cdot \bar \ell_{\not 0}) (s\cdot \bar \ell_{\not0}) }{d_p^2} \left[ \frac{4}{3} (s\cdot  \hat r_p) +   \frac{(\hat  r_p \cdot \bar \ell_{\not 0}) (s\cdot \bar \ell_{\not0}) }{2}\right]+ \frac{(s\cdot \bar \ell_{\not0})^2 r^2_p}{6 d^2_p}\right\}  \\
\fl  + \frac{\bar \ell^k_{\not0}}{3d^2_p} \left\{  -2(\hat  r_p \cdot \bar \ell_{\not 0})-(s\cdot  \hat r_p)^2 \hat r^2_p \frac{(s\cdot \bar \ell_{\not0})(s\cdot  \hat r_p)}{2} +\frac{( \hat r_p \cdot \bar \ell_{\not 0})^2  (s\cdot \bar \ell_{\not0})}{2} \left(-2(s\cdot  \hat r_p)  - ( \hat r_p \cdot d_s) \right)\right\} \\
\fl  - D_s^k  (s\cdot \hat  r_p);   
\end{eqnarray*}
%\hline
\begin{eqnarray*}
\fl C_6^k= \frac{D^{k}_{p}}{2}  (s \cdot \hat r_p)^2;
\end{eqnarray*}
%\hline
\begin{eqnarray*}
\fl C_7^k= -\frac{\bar \ell^k_{\not0}}{2} +   (s \cdot \bar \ell_{\not0})D^{k}_{s}; 
 \end{eqnarray*}
 %\hline
\begin{eqnarray*}
\fl C_8^k=   - \frac{D^{k}_{p}}{2}(s\cdot \hat r_p)^2 ( \hat r_p \cdot \bar \ell_{\not0})- \frac{\bar \ell^k_{\not0}}{2} \left\{(s \cdot \hat r_p)( \hat r_p \cdot \bar \ell_{\not0})  +(s\cdot \bar \ell_{\not0})^2 \left[ d_p^2 -  (\hat r_p \cdot \bar \ell_{\not 0})^2 \right] \right\};
\end{eqnarray*}
%\hline
\begin{eqnarray*}
\fl C_{9}^k= \frac{3D^{k}_{p}}{4d^3_p} \left\{ - [9 (s\cdot \hat r_p)(s\cdot \tilde v) + 2 (\hat r_p \cdot \tilde v)] + \frac{45}{2d^2_p} (s\cdot  \hat r_p)^2 (\hat r_p \cdot d_v) 
+  (\hat  r_p \cdot \bar \ell_{\not 0}) [9 (s\cdot \bar \ell_{\not0})(s\cdot \tilde v)  \right. \\
\fl  \left. + 2 (\bar \ell_{\not0} \cdot \tilde v)] - \frac{45}{d^2_p} (\hat r_p \cdot \bar \ell_{\not 0})(s\cdot \bar \ell_{\not0})(\hat r_p \cdot \tilde v)  (s\cdot \hat  r_p)+\frac{9}{2d^2_p} (s\cdot \bar \ell_{\not0})[4 ( \hat r_p \cdot \bar \ell_{\not 0})^2 + \hat r^2_p]  [2 (s\cdot  \hat r_p) (\bar \ell_{\not0} \cdot \tilde v)   \right.  \\
\fl \left.+ (s\cdot \bar \ell_{\not0}) (\hat r_p \cdot \tilde v) ]  + \frac{9}{2d^2_p}  [2 ( \hat r_p \cdot \bar \ell_{\not 0})^2 + 3 \hat r^2_p] (s\cdot \bar \ell_{\not0})^2 ( \hat r_p \cdot \bar \ell_{\not 0}) ( \ell_{\not0} \cdot \tilde v) ] \right\} \\
\fl + \frac{3D_s^k}{d_p} \left\{ (s\cdot \tilde v)- \frac{3(\hat r_p \cdot \tilde v)}{4d^2_p} (d_s\cdot \hat r_p)   + \frac{( \ell_{\not0} \cdot \tilde v) }{4d^2_p} \left[ 3(s\cdot  \hat r_p)( \hat r_p \cdot \bar \ell_{\not 0}) -   (s\cdot \bar \ell_{\not0}) [2(\hat  r_p \cdot \bar \ell_{\not 0})^2 + \hat r^2_p] \right] \right\}    \\
\fl  + \frac{3\bar \ell^k_{\not0}}{4d^3_p}\left\{ - ( \hat r_p \cdot \bar \ell_{\not 0}) [9 (s\cdot  \hat r_p)(s\cdot \tilde v) + 2 (\hat r_p \cdot \tilde v)]  + \frac{45}{2d^2_p} (\hat r_p \cdot \bar \ell_{\not 0})(s\cdot \hat  r_p)^2 (\hat r_p \cdot \tilde v)+  \frac{[2 ( \hat r_p \cdot \bar \ell_{\not 0})^2 +\hat  r^2_p]}{3} \right.  \\
\fl  \left.[9 (s\cdot \tilde v) (s\cdot \bar \ell_{\not0}) +2 (\bar \ell_{\not0} \cdot \tilde v) ] -  \frac{9[4 (\hat r_p \cdot \bar \ell_{\not 0})^2 + \hat r^2_p] }{2d^2_p}  [(\bar \ell_{\not0} \cdot \tilde v) \left( (s\cdot  \hat r_p)^2 
- \hat r^2_p (s\cdot \bar \ell_{\not0})^2 \right) + 2 (s \cdot \hat r_p)\right.  \\
\fl  \left. (s\cdot \bar \ell_{\not0}) (\hat r_p \cdot \tilde v)  ]- \frac{9[2 ( \hat r_p \cdot \bar \ell_{\not 0})^2 + 3\hat r^2_p] }{2d^2_p} (\hat r_p \cdot \bar \ell_{\not 0})(s\cdot \bar \ell_{\not0}) [2 (s \cdot \hat r_p) (\bar \ell_{\not0} \cdot \tilde v) + (s\cdot \bar \ell_{\not0}) (\hat r_p \cdot \tilde v)] \right \}   \\
\fl  + \frac{D^{k}_{v}}{2d_p} \left\{  1+   \frac{9(s \cdot \hat r_p)^2}{4d_p^2} +    \frac{9(s\cdot \bar \ell_{\not0})(s \cdot \hat r_p)(\hat  r_p \cdot \bar \ell_{\not 0})}{2d_p^2} + \frac{(s\cdot \bar \ell_{\not0})^2[2 (\hat r_p \cdot \bar \ell_{\not 0})^2 + \hat r^2_p] }{4d_p^2} \right\};
\end{eqnarray*}
%\hline
 \begin{eqnarray*}
 \fl C_{10}^k=- 27 D^{k}_{p}  \left\{ (s\cdot \bar \ell_{\not0})^2 (\bar \ell_{\not0} \cdot \tilde v)\left[\frac{( \hat r_p \cdot \bar \ell_{\not 0})^4}{4d^2_p} + \frac{2 \hat r^2_p + ( \hat r_p \cdot \bar \ell_{\not 0})^2}{4}\right] \right\}+  \frac{27 \bar \ell^k_{\not0}} {4}\left\{(s\cdot \bar \ell_{\not0}) \right.  \\
\fl \left.  \left[\frac{( \hat r_p \cdot \bar \ell_{\not 0})^4}{d^2_p} + 2 \hat r^2_p + (\hat r_p \cdot \bar \ell_{\not 0})^2\right]  [ 2 (s\cdot \hat  r_p)(\bar \ell_{\not0} \cdot \tilde v)+ (s\cdot \bar \ell_{\not0})(\hat r_p \cdot \tilde v)]  + (s\cdot \bar \ell_{\not0})^2 (\bar \ell_{\not0} \cdot \tilde v) \right. \\
\fl \left. \frac{17 (\hat r_p \cdot \bar \ell_{\not 0}) \hat r_p^4- 28\hat  r^2_p (\hat  r_p \cdot \bar \ell_{\not 0})^3+ 16 (\hat  r_p \cdot \bar \ell_{\not 0})^5  }{3d^2_p} \right\}; 
\end{eqnarray*}
%\hline
 \begin{eqnarray*}
\fl C_{11}^k= -D^{k}_{p}  \left\{ (s\cdot \bar \ell_{\not0})^2 (\bar \ell_{\not0} \cdot \tilde v) \frac{( \hat r_p \cdot \bar \ell_{\not 0})\hat r_p^2}{2d^2_p} \right\} +  \frac{3 d_p^2 D_s^k}{2} (\bar \ell_{\not0} \cdot \tilde v) (s\cdot \bar \ell_{\not0})  - \frac{d_p^2 D_v^k }{2}(s\cdot \bar \ell_{\not0})^2 \\
\fl + \bar \ell^k_{\not0}\left\{ d_p^2 [ \frac{9}{2} (s\cdot \bar \ell_{\not0}) (s\cdot \tilde v) + (\bar \ell_{\not0} \cdot \tilde v)] + \frac{27 (s\cdot \bar \ell_{\not0})}{4 d^2_p} \left[ \hat r_p^2 ( \hat r_p \cdot \bar \ell_{\not 0})  [2(\bar \ell_{\not0} \cdot \tilde v) (s\cdot \hat  r_p)+  (s\cdot \bar \ell_{\not0}) \right. \right. \\
 \fl \left. \left. (\hat r_p \cdot \tilde v)] + 9 (s\cdot \bar \ell_{\not0}) (\bar \ell_{\not0} \cdot \tilde v)  [ 7  \hat r_p^4- 20 \hat r^2_p ( \hat r_p \cdot \bar \ell_{\not 0})^2- 8 ( \hat r_p \cdot \bar \ell_{\not 0})^4] \right]\right\};
 \end{eqnarray*}
 %\hline
 \begin{eqnarray*} 
 \fl C_{12}^k= 9D^{k}_{p} d_p^2 \hat r^2_p (s\cdot \bar \ell_{\not0})^2 (\bar \ell_{\not0} \cdot \tilde v) +  9 \bar \ell^k_{\not0} 
 \left\{(s\cdot \bar \ell_{\not0})^2 (\bar \ell_{\not0} \cdot \tilde v) \hat r_p^2 (\hat  r_p \cdot \bar \ell_{\not 0})^3  - d^2_p \hat r^2_p  (s\cdot \bar \ell_{\not0}) \right. \\
 \fl \left. [ 2 (s\cdot  \hat r_p) (\bar \ell_{\not0} \cdot \tilde v)+ (s\cdot \bar \ell_{\not0})(\hat r_p\cdot \tilde v)]\right \}; 
\end{eqnarray*}
%\hline
 \begin{eqnarray*} 
\fl C_{13}^k= 9 D^{k}_{p} d_p^2  (s\cdot \bar \ell_{\not0}) [ 2 (s\cdot \hat  r_p) (\bar \ell_{\not0} \cdot \tilde v)+ (s\cdot \bar \ell_{\not0})(\hat r_p\cdot \tilde v)+ 3 (\hat r_p \cdot \bar \ell_{\not 0})(s\cdot \bar \ell_{\not0}) (\bar \ell_{\not0} \cdot \tilde v)] \\
 \fl +  9  \bar \ell^k_{\not0} \left\{(s\cdot \bar \ell_{\not0})^2 (\bar \ell_{\not0} \cdot \tilde v)   [\hat r^4_p- 5 d^2_p (\hat r_p \cdot \bar \ell_{\not 0})] - 3 d_p^2 (s\cdot \bar \ell_{\not0}) (\hat r_p \cdot \bar \ell_{\not 0}) [   2 (s\cdot \hat r_p) (\bar \ell_{\not0} \cdot \tilde v) \right. \\
 \fl \left.+ (s\cdot \bar \ell_{\not0})(\hat r_p\cdot \tilde v)]  - (s\cdot  \hat r_p)  d^2_p [  (s\cdot \hat  r_p) (\bar \ell_{\not0} \cdot \tilde v) +2 (\hat r_p\cdot \tilde v)(s\cdot \bar \ell_{\not0})  ]\right\}; 
\end{eqnarray*}
%\hline
 \begin{eqnarray*}
\fl C_{14}^k= \frac{3D^{k}_{p}}{2d^2_p} \left\{ 9 (s\cdot  \hat r_p)(s\cdot \tilde v) + 2 (\hat r_p \cdot \tilde v) + \frac{45}{2d^2_p} (s\cdot \hat r_p)^2 (\hat r_p \cdot d_v)  + ( \hat r_p \cdot \bar \ell_{\not 0}) [9 (s\cdot \bar \ell_{\not0})(s\cdot \tilde v)  \right. \\
\fl  \left. + 2 (\bar \ell_{\not0} \cdot \tilde v)] - \frac{45}{d^2_p} (\hat r_p \cdot \bar \ell_{\not 0})(s\cdot \bar \ell_{\not0})(\hat r_p \cdot \tilde v) (s\cdot  \hat r_p)  + \frac{9}{2d^2_p} \left[4 (\hat r_p \cdot \bar \ell_{\not 0})^2 + \hat r^2_p - (s\cdot \bar \ell_{\not0})(\hat  r_p \cdot \bar \ell_{\not 0})  \right. \right. \\
\fl \left. \left. \left( (\hat r_p \cdot \bar \ell_{\not 0})^2 - \frac{3 \hat r^2_p}{2} \right) \right] [2 (s\cdot  \hat r_p) (\bar \ell_{\not0} \cdot \tilde v)+ (s\cdot \bar \ell_{\not0}) (\hat r_p \cdot \tilde v) ]  \right\} + 3D_s^k \left\{ (s\cdot \tilde v)+ \frac{(\hat  r_p \cdot \bar \ell_{\not 0})}{d^2_p} \right.  \\
\fl \left.  [(\hat r_p \cdot \tilde v )(s\cdot \bar \ell_{\not0})  + (s\cdot \hat r_p)(\bar \ell_{\not0} \cdot \tilde v)  ]-  \frac{( \ell_{\not0} \cdot \tilde v)}{4d^2_p} [2(\hat  r_p \cdot \bar \ell_{\not 0})^2 + \hat r^2_p] (s\cdot \bar \ell_{\not0}) \right\} + \frac{3\bar \ell^k_{\not0}}{4d^2_p} \left\{ - (\hat r_p \cdot \bar \ell_{\not 0})  \right.\\
\fl  \left. [9 (s\cdot  \hat r_p)(s\cdot \tilde v)+ 2 (\hat r_p \cdot \tilde v)] + \frac{45}{2d^2_p} ( r_p \cdot \bar \ell_{\not 0})(s\cdot \hat r_p)^2 (\hat r_p \cdot \tilde v)  + [2 (\hat  r_p \cdot \bar \ell_{\not 0})^2 + \hat r^2_p]  [\frac{3}{2}(s\cdot \tilde v) (s\cdot \bar \ell_{\not0})   \right.  \\
\fl \left. + \frac{1}{3} (\bar \ell_{\not0} \cdot \tilde v) ] -  \frac{9[4 (\hat r_p \cdot \bar \ell_{\not 0})^2 + \hat r^2_p] }{2d^2_p} [(\bar \ell_{\not0} \cdot \tilde v) \left( (s\cdot \hat r_p)^2 - \hat r^2_p (s\cdot \bar \ell_{\not0})^2 \right) + 2 (s \cdot \hat r_p)(s\cdot \bar \ell_{\not0})(\hat r_p \cdot \tilde v)]  \right.\\
\fl \left. -  \frac{9(\hat  r_p \cdot \bar \ell_{\not 0}) (s \cdot \bar \ell_{\not0} )[2 ( \hat r_p \cdot \bar \ell_{\not 0})^2 +3 \hat r^2_p] }{2d^2_p} [2 (\bar \ell_{\not0} \cdot \tilde v)  (s\cdot \hat r_p) + (s\cdot \bar \ell_{\not0}) (\hat r_p \cdot \tilde v)  ] \right\} \\
\fl  + \frac{D^{k}_{v}}{2} \left\{  1-  \frac{(s \cdot \hat r_p)^2}{d_p^2} +    \frac{9(s\cdot \bar \ell_{\not0})(s \cdot \hat r_p)(\hat r_p \cdot \bar \ell_{\not 0})}{2d_p^2}  + \frac{(s\cdot \bar \ell_{\not0})^2 [2 (\hat  r_p \cdot \bar \ell_{\not 0})^2 +\hat  r^2_p] }{4d_p^2} \right\};
\end{eqnarray*}
%\hline
\begin{eqnarray*}
\fl C_{15}^k= \frac{D^{k}_{p}}{2} \left\{  (s\cdot \hat r_p)(s\cdot \tilde v) + \frac{2}{9} (\hat r_p \cdot \tilde v) + \frac{5}{d^2_p} (s\cdot \hat r_p)  \left[(s\cdot \hat  r_p)  (\hat r_p \cdot \tilde v) -  ( \hat r_p \cdot \bar \ell_{\not 0})\left( (\bar \ell_{\not0} \cdot \tilde v)  \right.  \right. \right.\\
\fl \left.\left. \left. (s\cdot \hat r_p) +2 (s\cdot \bar \ell_{\not0}) (\hat r_p \cdot \tilde v)\right) \right]+ 3 (s\cdot \bar \ell_{\not0})\frac{4 ( \hat r_p \cdot \bar \ell_{\not 0})^2 +\hat r^2_p}{2d^2_p}   [2 (s\cdot \hat r_p) (\bar \ell_{\not0} \cdot \tilde v)+ (s\cdot \bar \ell_{\not0}) (\hat r_p \cdot \tilde v) ] \right\}   \\
\fl - \frac{3D^{k}_{v}}{4}  (s \cdot \hat r_p)^2 + \frac{3D_s^k}{2}  (s\cdot \hat  r_p)  (\hat r_p \cdot \tilde v)  + \frac{9\bar \ell^k_{\not0}}{4d^2_p} \left\{5 (\hat r_p \cdot \bar \ell_{\not 0})(s\cdot  \hat r_p)^2 (\hat r_p \cdot \tilde v) \right. \\
\fl  \left. - 3(s\cdot  \hat r_p)  [4 (\hat  r_p \cdot \bar \ell_{\not 0})^2 + \hat r^2_p] [(\bar \ell_{\not0} \cdot \tilde v)  (s\cdot  \hat r_p) + 2(s\cdot \bar \ell_{\not0}) (\hat r_p \cdot \tilde v)  ]\right\}; 
\end{eqnarray*}
%\hline
\begin{eqnarray*}
\fl C_{16}^k=9D^{k}_{p}  (s\cdot  \hat r_p)^2 (\hat r_p \cdot \tilde v); 
\end{eqnarray*}
%\hline
\begin{eqnarray*}
\fl C_{17}^k= \frac{D^{k}_{p}}{2} \left\{ [9(s\cdot \bar \ell_{\not0}) (s\cdot \tilde v)+ 2 (\bar \ell_{\not0} \cdot \tilde v) ]  \right\} +\frac{\bar \ell^k_{\not0}}{2}\left\{( \hat r_p \cdot \bar \ell_{\not 0})  [9(s\cdot \bar \ell_{\not0})(s\cdot  \tilde v) + 2 (\bar \ell_{\not0} \cdot \tilde v)] \right. \\
\fl \left. - [9(s\cdot \hat  r_p)(s\cdot \tilde v)  + 2 (\hat r_p \cdot \tilde v)] \right \}  + \frac{3D_s^k}{2} [ (s\cdot \bar \ell_{\not0}) (\hat r_p \cdot \tilde v) +  (\bar \ell_{\not0} \cdot \tilde v) \left((s\cdot \hat r_p)  \right. \\
\fl \left. - (s\cdot \bar \ell_{\not0})(\hat r_p \cdot \bar \ell_{\not 0}) \right)]+ \frac{3D_v^k}{4}(s\cdot \bar \ell_{\not0})^2  ( \hat r_p \cdot \bar \ell_{\not 0});
\end{eqnarray*}
%\hline
\begin{eqnarray*}
\fl C_{18}^k=- 9D^{k}_{p} \left\{ 3 (s\cdot \hat r_p)[(s\cdot \hat r_p)(\bar \ell_{\not0} \cdot \tilde v)+ 2 (s\cdot \bar \ell_{\not0}) (\hat r_p \cdot \tilde v)]  - (s\cdot \bar \ell_{\not0})( \hat r_p \cdot \bar \ell_{\not 0})[2(s\cdot \hat r_p)(\bar \ell_{\not0} \cdot \tilde v)  \right. \\
\fl \left. +  (s\cdot \bar \ell_{\not0}) (\hat r_p \cdot \tilde v)] - 3 (s\cdot \bar \ell_{\not0})^2 (\bar \ell_{\not0} \cdot \tilde v)( \hat r_p \cdot \bar \ell_{\not 0})^2 \right\} - 9\bar \ell^k_{\not0} \left\{3 (s\cdot  \hat r_p)^2  (\hat r_p \cdot \tilde v) + (s\cdot \hat r_p)(\hat r_p \cdot \bar \ell_{\not 0})  \right. \\
\fl  \left. [(s\cdot \hat r_p)(\bar \ell_{\not0} \cdot \tilde v)+ 2 (s\cdot \bar \ell_{\not0}) (\hat r_p \cdot \tilde v)] +  (s\cdot \bar \ell_{\not0}) (\hat r_p \cdot \bar \ell_{\not 0})^2 [2(s\cdot \hat r_p)(\bar \ell_{\not0} \cdot \tilde v)+  (s\cdot \bar \ell_{\not0}) (\hat r_p \cdot \tilde v)  \right.\\ 
\fl \left. - d^2_p (s\cdot \bar \ell_{\not0})^2 ( \hat r_p \cdot \bar \ell_{\not 0}) (\bar \ell_{\not0} \cdot \tilde v) ] \right\}; 
\end{eqnarray*}
%\hline
\begin{eqnarray*}
\fl C_{19}= - \frac{1}{2d_p^2} \left\{ [2d_p^4 - 3 \hat r^2_p d_p^2 +3( \hat r_p \cdot \bar \ell_{\not 0})^4 )] (s\cdot \bar \ell_{\not0})^2(\bar \ell_{\not0} \cdot \tilde v)\right\};
\end{eqnarray*}
%\hline
\begin{eqnarray*} 
\fl C_{20}= - \frac{3\hat r^2_p}{2d_p^2}   (s\cdot \bar \ell_{\not0})^2(\bar \ell_{\not0} \cdot \tilde v)(\hat r_p \cdot \bar \ell_{\not 0});  
\end{eqnarray*}
%\hline
\begin{eqnarray*}
\fl C_{21}=  3d_p^2   (s\cdot \bar \ell_{\not0}) \left[ \frac{1}{2} (\hat r_p \cdot \tilde v) (s\cdot \bar \ell_{\not0}) + (\bar \ell_{\not0} \cdot \tilde v)  (\hat  r_p \cdot d_s) \right]; 
\end{eqnarray*}
%\hline
\begin{eqnarray*}
\fl C_{22} =   \frac{1}{2d_p^2} \left\{ (\hat r_p \cdot d_v) - \frac{4}{d^2_p} (\hat r_p \cdot \tilde v) (s\cdot \hat r_p)^2 + \frac{4}{d^2_p}  (s\cdot  \hat r_p) ( \hat r_p \cdot \bar \ell_{\not 0}) [(s\cdot  \hat r_p)(\bar \ell_{\not0} \cdot \tilde v) \right. \\
\fl \left.+2  (s\cdot \bar \ell_{\not0}) (\hat r_p \cdot \tilde v)]  -\frac{(s\cdot \bar \ell_{\not0})}{d^2_p}[\hat r^2_p + 3 ( \hat r_p \cdot \bar \ell_{\not 0})^2]  [2(s\cdot  \hat r_p)(\bar \ell_{\not0} \cdot \tilde v)+ (s\cdot \bar \ell_{\not0}) (\hat r_p \cdot \tilde v)]\right\}; 
\end{eqnarray*}
%\hline
\begin{eqnarray*}
\fl C_{23} =  \frac{1}{2d_p^2} \left\{ 3(\hat r_p \cdot \tilde v)d^2_p  - 4 (\hat r_p \cdot \tilde v) (s\cdot \hat  r_p)^2 + 4 (\hat  r_p \cdot \bar \ell_{\not 0})  (s\cdot  \hat r_p)[(s\cdot  \hat r_p)(\bar \ell_{\not0} \cdot \tilde v)+2  (s\cdot \bar \ell_{\not0})  \right. \\
\fl \left.(\hat r_p \cdot \tilde v)] +(s\cdot \bar \ell_{\not0}) [\hat r^2_p + 3 (\hat r_p \cdot \bar \ell_{\not 0})^2][2(s\cdot  \hat r_p)(\bar \ell_{\not0} \cdot \tilde v)+  (\hat r_p \cdot \tilde v)(s\cdot \bar \ell_{\not0})] \right. \\
\fl \left. +(s\cdot \bar \ell_{\not0})^2  (\bar \ell_{\not0} \cdot \tilde v)(\hat  r_p \cdot \bar \ell_{\not 0})^3 \right\};
\end{eqnarray*}
%\hline
\begin{eqnarray*}
\fl C_{24}= (s\cdot \hat r_p)^2 (\hat r_p \cdot \tilde v); 
\end{eqnarray*}
%\hline
\begin{eqnarray*}
\fl C_{25}= -\frac{ (\bar \ell_{\not0} \cdot \tilde v)}{2};    
\end{eqnarray*}
%\hline
\begin{eqnarray*}
\fl C_{26}= - \frac{3}{2} \left\{  (s\cdot  \hat r_p)[(s\cdot \hat  r_p)(\bar \ell_{\not0} \cdot \tilde v)+2  (s\cdot \bar \ell_{\not0}) (\hat r_p \cdot \tilde v)]+   ( \hat r_p \cdot \bar \ell_{\not 0})  (s\cdot \bar \ell_{\not0}) [2(s\cdot  \hat r_p)(\bar \ell_{\not0} \cdot \tilde v) \right. \\
\fl  \left.+  (s\cdot \bar \ell_{\not0}) (\hat r_p \cdot \tilde v)] + (s\cdot \bar \ell_{\not0})^2  (\bar \ell_{\not0} \cdot \tilde v) [d^2_p - ( \hat r_p \cdot \bar \ell_{\not 0})^2] \right\}; 
\end{eqnarray*}
%\hline
\begin{eqnarray*}
\fl C_{27} =  \frac{(s\cdot \bar \ell_{\not0})^2}{3} [ 2d^4_p - 3\hat r^2_p d^2_p -3 ( \hat r_p \cdot \bar \ell_{\not 0})^4 ]; 
\end{eqnarray*}
%\hline
\begin{eqnarray*}
\fl C_{28} =  -\frac{\hat r^2_p}{d^2_p}  (s\cdot \bar \ell_{\not0})^2  (\hat r_p \cdot \bar \ell_{\not 0});
\end{eqnarray*}
%\hline
\begin{eqnarray*}
\fl C_{29}= d^2_p (s\cdot \bar \ell_{\not0}) [ ( \hat r_p \cdot \bar \ell_{\not 0})(s\cdot \bar \ell_{\not0}) + 2 (s\cdot \hat r_p) ]; 
\end{eqnarray*}
%\hline
\begin{eqnarray*}
\fl C_{30} = - \frac{4}{d^2_p}  (s\cdot \bar \ell_{\not0}) ( s\cdot d_p)  
-  \frac{2}{3d^4_p}   (\hat r_p \cdot \bar \ell_{\not 0}) \left [2 (s\cdot \hat r_p) \left((s\cdot \bar \ell_{\not0})(\hat r_p \cdot \bar \ell_{\not 0}) +  2 (s\cdot \hat r_p) \right) \right.\\
\fl  \left. -  (s\cdot \bar \ell_{\not0})^2 [\hat r^2_p +  (\hat  r_p \cdot \bar \ell_{\not 0})^2] \right];
\end{eqnarray*}
%\hline
\begin{eqnarray*}
\fl C_{31}=  -  ( \hat r_p \cdot \bar \ell_{\not 0}) - 6 (\hat  r_p \cdot s)(s\cdot \bar \ell_{\not0})+ \frac{4}{3d^2_p} ( \hat r_p \cdot s) ( \hat r_p \cdot \bar \ell_{\not 0}) [ 2 (s\cdot \bar \ell_{\not0})  (\hat  r_p \cdot \bar \ell_{\not 0})+ 3 (\hat r_p \cdot s) ] \\
\fl + \frac{\hat r^2_p +3(\hat r_p \cdot \bar \ell_{\not 0})^2  }{3d^2_p}  (s\cdot \bar \ell_{\not0})[( \hat r_p \cdot \bar \ell_{\not 0})  (s\cdot \bar \ell_{\not0})+ 2 ( \hat r_p \cdot s)] - \frac{(\hat r_p \cdot \bar \ell_{\not 0})^3  (s\cdot \bar \ell_{\not0})^2 }{3d^2_p}; 
\end{eqnarray*}
%\hline
\begin{eqnarray*}
\fl C_{32}=   (s\cdot \hat r_p)^2 (\hat r_p \cdot \bar \ell_{\not 0});
\end{eqnarray*}
%\hline
\begin{eqnarray*}
\fl C_{33}=  \frac{1 }{3}[1+ 6 (s \cdot \bar \ell_{\not0})^2].    
%&{}&\nonumber\\
%C_{34}&= & d^2_p ( s \cdot \bar \ell_{\not0})^2 - (s \cdot r_p)^2     \nonumber \\
%&{}&\nonumber\\
%C_{35}&= &  \frac{2}{d_p^2} [ (s\cdot r_p) + (  r_p \cdot \bar \ell_{\not0} ) (s \cdot \bar \ell_{\not0}) ]     \nonumber \\
%&{}&\nonumber\\
%C_{36}&= &  -3 (s \cdot r_p)   \nonumber \\
%&{}&\nonumber\\
%C_{37}&= &  (s \cdot \bar \ell_{\not0})  \nonumber 
\end{eqnarray*}

\end{document}